\documentclass[10pt, two column, twoside]{IEEEtran}
\usepackage{lipsum}
\usepackage{cuted}
\usepackage{amsthm}
\usepackage{amsmath}
\usepackage{algorithmic}
\usepackage{algorithm}
\usepackage{mathrsfs}
\usepackage{graphicx}
\usepackage{subfigure}
\usepackage{cite}
\usepackage{bm,epsfig,amsthm,url}
\usepackage{indentfirst}
\usepackage{amssymb}
\usepackage{epstopdf}
\usepackage{xcolor}
\usepackage{mathtools}

\theoremstyle{definition}

\newtheorem{remark}{Remark}

\begin{document}

\title{Multi-Passive/Active-IRS Enhanced Wireless Coverage:  Deployment Optimization and Cost-Performance Trade-off}
\author{Min~Fu,~\IEEEmembership{Member,~IEEE}, Weidong~Mei,~\IEEEmembership{Member,~IEEE}, and~Rui~Zhang,~\IEEEmembership{Fellow,~IEEE}\vspace{-2em}
	\thanks{Min Fu is with the Department of Electrical and Computer Engineering, National University of Singapore, Singapore 117583 (e-mail:  fumin@nus.edu.sg).}
		\thanks{W. Mei is with the National Key Laboratory of Wireless Communications, University of Electronic Science and Technology of China, Chengdu 611731, China (e-mail: wmei@uestc.edu.cn)} 
			\thanks{R. Zhang is with School of Science and Engineering, Shenzhen Research Institute of Big Data, The Chinese University of Hong Kong, Shenzhen, Guangdong 518172, China (e-mail: rzhang@cuhk.edu.cn). He is also with the Department of Electrical and Computer Engineering, National University of Singapore, Singapore 117583 (e-mail: elezhang@nus.edu.sg).}						
}

\maketitle

\setlength\abovedisplayskip{2pt}
\setlength\belowdisplayskip{2pt}
\setlength\abovedisplayshortskip{2pt}
\setlength\belowdisplayshortskip{2pt}
\setlength\arraycolsep{2pt}

\begin{abstract}

Both passive and active intelligent reflecting surfaces (IRSs) can be deployed in complex environments to enhance wireless network coverage by creating multiple blockage-free cascaded line-of-sight (LoS) links.
In this paper, we study a multi-passive/active-IRS (PIRS/AIRS) aided wireless network with a multi-antenna base station (BS) in a given region.
First, we divide the region into multiple non-overlapping cells, each of which may contain one candidate location that can be deployed with a single PIRS or AIRS.
Then, we show several trade-offs between minimizing the total IRS deployment cost and enhancing the signal-to-noise ratio (SNR) performance over all cells via direct/cascaded LoS transmission with the BS.
To reconcile these trade-offs, we formulate a joint multi-PIRS/AIRS deployment problem to select an optimal subset of all candidate locations for deploying IRS and also optimize the number of passive/active reflecting elements deployed at each selected location to satisfy a given SNR target over all cells, such that the total deployment cost is minimized. 
However, due to the combinatorial optimization involved, the formulated problem is difficult to be solved optimally. 
To tackle this difficulty, we first optimize the reflecting element numbers with given PIRS/AIRS deployed locations via sequential refinement, followed by a partial enumeration to determine the PIRS/AIRS locations.
Simulation results show that our proposed algorithm achieves better cost-performance trade-offs than other baseline deployment strategies.

\end{abstract}
\begingroup\allowdisplaybreaks

\section{Introduction}

Intelligent reflecting surface (IRS) has received increasingly high attention in wireless communications due to its passive, full-duplex, and controllable signal reflection, which can improve the spectral and energy efficiency of future wireless networks cost-effectively \cite{Wu2021Tutorial, Mei2022Multireflection}.
Specifically, IRS consists of a large array of passive reflecting elements, each of which can be dynamically tuned to alter the amplitude/phase of its reflected signal \cite{Wu2021Tutorial}.
However, IRS incurs multiplicative path loss over its cascaded channel, which may limit the signal coverage performance.
To compensate for the multiplicative path loss, a new active IRS (AIRS) architecture has recently been proposed, where each reflecting element is equipped with an active load (or called negative resistance), such that it can reflect the incident signal with additional power amplification \cite{zhang2022active, Long2021active, Chen2023active, Fu2022active,You2021AIRS}.
However, compared to the conventional passive IRS (PIRS), the AIRS induces higher hardware cost and non-negligible amplification noise in its reflected signals, which may degrade the communication performance especially when the number of reflecting elements is large and/or its amplification gain is low \cite{You2021AIRS}. 
As such, the joint use of PIRS and AIRS emerges as an appealing solution to reap their complementary advantages \cite{kang2023active}.
\begin{figure}[t]
	\centering
	\includegraphics[scale=0.3]{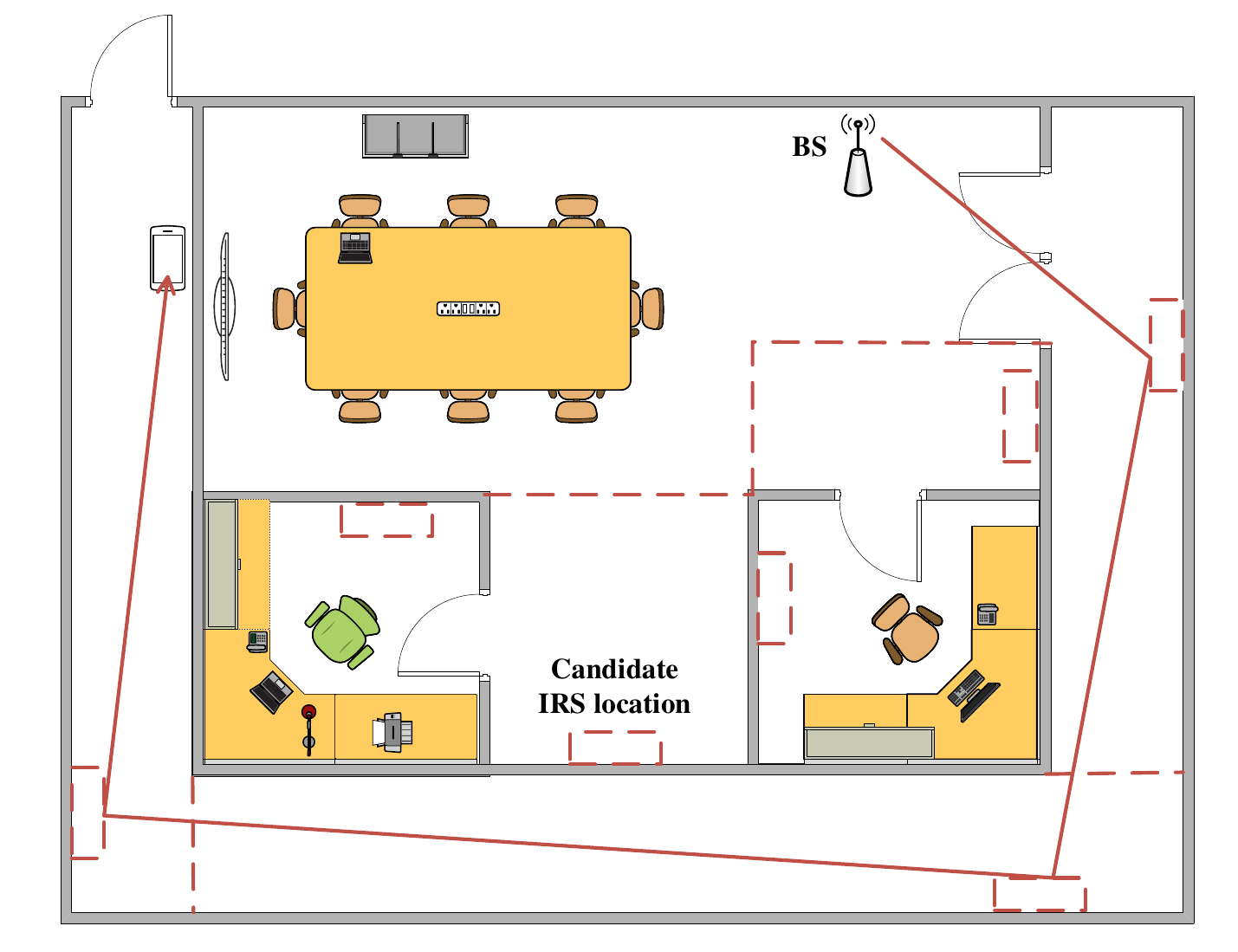}
	\vspace{-2mm}		
	\caption{Multi-IRS aided wireless network in a typical indoor environment.} \label{Fig:systemmodel0}
	\vspace{-6mm}
\end{figure}

Furthermore, to practically reap the beamforming gain by PIRS/AIRS, their deployment should be carefully designed to ensure the reflected signal coverage considering their half-space reflection constraints as well as the practical environment conditions.
Therefore, there have been some prior works addressing the deployment optimization for PIRS/AIRS.
For example, in the case of PIRS, the authors in \cite{Mu2021SingleReflection} formulated and solved a PIRS location optimization problem to maximize the weighted sum rate in multi-user communications under three different multiple access schemes.
Moreover, the authors in \cite{Bai2022SingleReflection} optimized the PIRS's location in a secure wireless communication system to maximize the secrecy rate.
A deployment optimization problem for an unmanned aerial vehicle (UAV)-mounted PIRS was formulated and solved in \cite{Lu2021SingleReflect} to maximize the worst-case signal-to-noise ratio (SNR) among all user locations in a target area.
In addition to the PIRS's location optimization, the authors in \cite{Zeng2021Orientation} and \cite{ Cheng2022Aerial} further optimized its rotation and showed their joint effectiveness in terms of performance enhancement.
Furthermore, the authors in \cite{Hashida2020PIRS, Huang2022multiplePIRS, Huang2022cost, Efrem2023IRSdeploy} delved into the multi-PIRS deployment design in a given area and aimed to optimize the locations and/or number of PIRSs deployed.
In the case of AIRS, the authors in \cite{You2021AIRS} and \cite{kang2023active} aimed to optimize the location of an AIRS to maximize the achievable rate in a single-user system, which revealed that the optimal location of an AIRS generally differs from that of a PIRS.

All the above works, however, focused only on the deployment design for either PIRS or AIRS, rather than their joint use.
Furthermore, they took into account only the single reflection by a PIRS/an AIRS, while the multi-IRS reflection may also be exploited to create blockage-free cascaded line-of-sight (LoS) signal paths between multiple base stations (BSs) and distributed user locations, which helps enhance the wireless coverage performance particularly in complex environments with dense and scattered obstacles (e.g., the indoor environment shown in Fig. \ref{Fig:systemmodel0}).
Moreover, the severe multiplicative path loss due to multi-IRS signal reflections can be compensated for by the AIRS amplification gain and/or the pronounced cooperative passive beamforming (CPB) gain via successive PIRS reflections \cite{Mei2022Multireflection}.
Due to the above benefits, the authors in \cite{Mei2021routing} and \cite{Mei2022MIMO} studied a new beam routing problem in a multi-PIRS-reflection aided wireless network, which aims to select an optimal multi-PIRS-reflection path from a BS to each user and optimize the beamforming at the BS and selected IRSs, such that the received SNR at each user is maximized.
Such a beam routing problem was later extended to the case with joint use of an AIRS and multiple PIRSs in \cite{Zhang2023MAMP}. However, the above works \cite{Mei2021routing, Mei2022MIMO,Zhang2023MAMP} assumed known IRSs' locations without their deployment optimization. 
While our recent work \cite{Fu2023multi} considered a given multi-IRS reflection path formed by one AIRS and multiple PIRSs, and optimized the AIRS's location over this path jointly with all IRSs' beamforming designs to maximize the received SNR at the user.
However, this work only considered a single AIRS and assumed fixed PIRS locations.
Thus, it remains unknown how to properly deploy multiple PIRSs and AIRSs in a general multi-IRS-reflection aided wireless network.
It is worth noting that in \cite{Mei2023IRSdeploy}, we have studied a relevant multi-IRS deployment problem to jointly optimize the locations of multiple PIRSs in a multi-PIRS-reflection aided wireless network.
Nevertheless, its results cannot be directly applied to the case with joint use of PIRSs and AIRSs due to their different signal reflection models.
In addition, the communication performance in \cite{Mei2023IRSdeploy} was evaluated in terms of the number of signal reflections per multi-PIRS link, which may not accurately indicate the SNR performance of each user.

To tackle the above issues, in this paper, we investigate a new multi-IRS deployment problem in a general multi-PIRS/AIRS-reflection-aided wireless network.
The main contributions of this paper are summarized as follows:
\begin{itemize}
	\item First, we propose a graph-based system model for the joint PIRS and AIRS deployment by dividing the considered region into multiple non-overlapping cells, each of which may contain one predetermined candidate location for deploying a single PIRS or AIRS with a maximum allowable size, as shown in Fig. \ref{Fig:systemmodel0}.
	By this means, we then characterize the total cost of any given IRS deployment design and its resulting communication performance for each cell, which is measured by the maximum worst-case received SNR among all desired user locations within this cell over its all possible direct and cascaded LoS links with the BS. 
	It is also shown that there exist several cost-performance trade-offs based on the above characterization.
	
	\item Next, to optimally resolve these trade-offs, we formulate a joint PIRS and AIRS deployment problem, which aims at selecting an optimal subset of all candidate locations to deploy PIRSs/AIRSs and optimizing the number of reflecting elements deployed at each selected location, so that a given SNR target can be met for all cells while minimizing the total deployment cost.
	However, such a joint deployment problem is NP-hard and generally difficult to be solved optimally.
	To address this challenge, we first optimize the number of reflecting elements per candidate location for given PIRS/AIRS deployed locations using the sequential refinement method. Then, based on the obtained solution, we propose a partial enumeration method to determine the PIRS/AIRS deployed locations.
	Numerical results show that our proposed algorithm can achieve near-optimal performance of full enumeration and yield better cost-performance trade-offs than other baseline deployment strategies.
	
\end{itemize}


The remainder of this paper is organized as follows.
Section II describes the system model.
Section III characterizes the total IRS deployment cost and SNR performance in the considered system.
Section IV presents the problem formulation for the joint PIRS and AIRS deployment design.
Section V presents our proposed algorithms to solve this problem.
Section VI presents numerical results to show the effectiveness of our proposed algorithms.
Section VII concludes this paper and discusses future work directions.

\textit{Notations}:
$\tbinom{n}{k}$ denotes the number of combinations to choose $k$ elements from a set of $n$ elements.
$\mathbb{N}^+$ denotes the set of positive integers.
$\jmath$ denotes the imaginary unit.
$x\sim \mathcal{CN}(0, \sigma^2)$ 
represents a circularly symmetric
complex Gaussian random variable $x$ with zero mean and variance $\sigma^2$.
$\mathbb{E}(\cdot)$ denotes the statistical expectation.
$(\cdot)^{\sf H}$ and $(\cdot)^{\sf T}$ denote the conjugate transpose and transpose, respectively.
For a complex-valued vector $\bm x$, $\|\bm x\|$ denotes its Euclidean norm, and $\text{diag}(\bm x)$ denotes a diagonal matrix with each diagonal entry being the corresponding element in $\bm x$.
For a set $\cal{S}$, $|\cal{S}|$ denotes its cardinality. $\emptyset$ denotes an empty set.
For two sets $\cal{S}$ and $\cal{S}'$, $\cal{S}\cap \cal{S}'$ denotes their intersection, $\cal{S}\cup \cal{S}'$ denotes their union, and $\cal{S}\backslash \cal{S}'$ is the set of elements that belong to $\cal{S}$ but are not in $\cal{S}'$.

\begin{figure*}[t]
	\centering
	\includegraphics[scale=0.25]{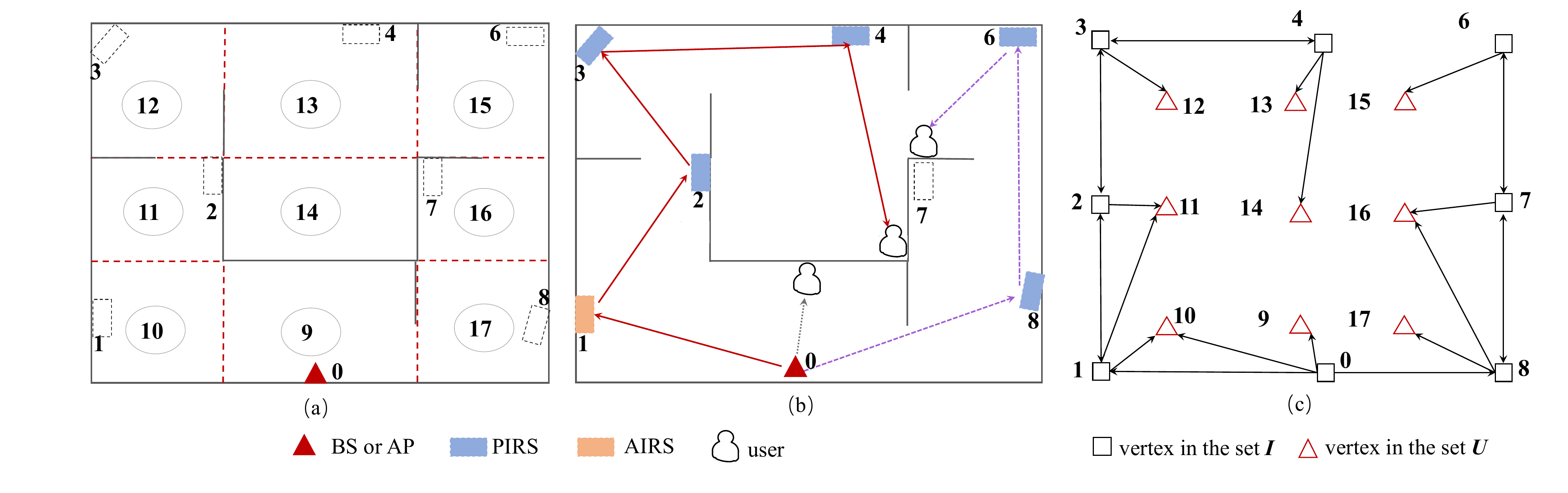}
	\vspace{-4mm}		
	\caption{Illustrations for (a) the cells in region $\cal D$ and the nodes representing the BS or candidate IRS locations,  (b) LoS paths created by PIRS and/or AIRS deployed at partial candidate locations, and (c) the corresponding graph $G$ of considered region ${\cal D}$.} \label{Fig:systemmodel}
	\vspace{-4mm}
\end{figure*}

\section{System Model}
As shown in Fig. \ref{Fig:systemmodel0}, in this paper, we study a wireless communication system in a given region, denoted by ${\cal D}$, where dense obstacles severely block a large portion of communication links.
Assume that a single BS (or access point (AP)) equipped with $M$ antennas has already been deployed in ${\cal D}$ to establish direct LoS links with as many desired user locations as possible.
To further boost network coverage, we consider that multiple IRSs, including PIRSs and AIRSs, can be deployed in $\cal D$ to create virtual LoS paths from the BS to other desired user locations in ${\cal D}$.
To facilitate their deployment, we assume that a number of candidate locations, denoted by $I_0$, have been identified in $\cal D$, each of which may be deployed with either a PIRS or an AIRS.
Furthermore, it is assumed that deploying IRSs at all these candidate locations would enable achieving global LoS coverage from the BS to any user location in $\cal D$; however, such a deployment incurs practically prohibitive deployment cost.
Therefore, due to the substantial deployment cost, only a subset of these candidate locations may be selected for deploying IRSs depending on prescribed coverage and communication performance requirements, as pursued in this paper.

To ease IRS deployment, similarly to \cite{Mei2023IRSdeploy}, we divide $\cal D$ into $J$ ($J\geq I_0$) non-overlapping cells, where the BS is deployed in cell $0$, and each of the remaining cells contains at most one candidate IRS location, as shown in Fig. \ref{Fig:systemmodel}(a).
Additionally, we assume that LoS coverage can be locally achieved between the candidate IRS location (or the BS) and any possible user locations in its located cell.
Let ${\cal J} \triangleq \{0,\ldots, J-1\}$ denote the set of all cells and ${\cal I}_0$ denote the set of cells containing candidate IRS locations, with ${\cal I}_0\subseteq {\cal J}$ and $|{\cal I}_0| = I_0$.
Moreover, we denote by ${\cal P}$ and ${\cal A}$ the sets of cells deployed with PIRSs and AIRSs, respectively, with ${\cal P} \subseteq {\cal I}_0, {\cal A} \subseteq {\cal I}_0$, and ${\cal P} \cap {\cal A} = \emptyset$.
For convenience, we refer to the IRS deployed in cell $i,i\in {\cal I}_0$, as IRS $i$.
In particular, it is also referred to as PIRS $i$ or AIRS $i$ if $i\in {\cal P}$ or $i\in {\cal A}$, respectively.

Furthermore, to mount the IRS efficiently at each candidate location, we consider that each IRS is assembled by 
a number of tiles (or subsurfaces) of the same fixed size when mounting it.
Let $N$ denote the number of reflecting elements in both horizontal and vertical dimensions per tile and $T_{i}$ denote the number of tiles on IRS $i$; hence, its total number of reflecting elements is given by $N_i \triangleq T_iN^2$.
Note that due to the practically limited size of IRS deployment, we assume that there is a maximum allowable number of tiles that can be deployed at each candidate location, denoted as $T^{\max}_0$. As such, we have $T_i \le T^{\max}_0, i \in {\cal I}_0$.
For each cell $i$ without IRS deployed (i.e., in the set ${\cal I}_0\backslash ({\cal P}\cup {\cal A})$), we set $T_i=0$.
For convenience, in the sequel of this paper, we refer to tile numbers as those in the cells deployed with IRSs, i.e., ${\cal P}\cup {\cal A}$. Accordingly, let ${\cal T} \triangleq \{T_i|i\in {\cal P}\cup {\cal A}\}$ denote the ensemble of numbers of tiles deployed in these cells.

For PIRS $p, p\in {\cal P}$, let ${\bm \Phi}_p={\rm diag}\{e^{\jmath\theta_{p,1}},\ldots,e^{\jmath\theta_{p,N_{p}}}\} \in {\mathbb C}^{N_{p} \times N_{p}}$ denote its reflection coefficient matrix, where
$\theta_{p,n}\in [0,2\pi]$ denotes the phase shift of the $n$-th reflecting element, and its amplitude is set to one to maximize the reflected signal power \cite{Wu2021Tutorial}.
While for AIRS $a, a\in {\cal A}$, we denote its reflection coefficient matrix as ${\bm \Phi}_a= {\rm diag}\{\eta_a e^{\jmath\theta_{a,1}},\ldots,\eta_a e^{\jmath\theta_{a, N_{a}}}\} \in {\mathbb C}^{N_{a} \times N_{a}}$, where $\eta_a$ ($\eta_a >1$) denotes a common amplification factor for all of its reflecting elements.
Unlike the PIRSs, each AIRS introduces non-negligible amplification noise into its reflected signal.
For AIRS $a, a\in {\cal A}$, we denote its amplification noise by $\bm{n}_{a}\in\mathbb{C}^{N_{a}\times 1}$, where $\bm{n}_{a}\sim \mathcal{CN}(\boldsymbol{0}_{N_{a}}, \sigma^2{\bf I}_{N_{a}})$ with $\sigma^2$ denoting the noise power.

For convenience, we refer to the BS and the candidate IRS location in cell $i, i\in {\cal I}_0$ as nodes 0 and $i$, respectively.
To describe the LoS availability between any two nodes in the network, we define a set of binary variables $\mu_{i, i'}$, $i\neq i', i, i' \in {\cal I}\triangleq\{0\}\cup {\cal I}_0$ to indicate the LoS availability between nodes $i$ and $i'$ (by setting $\mu_{i, i} = 0$).
In particular, $\mu_{i, i'}=1$ holds if and only if (iff) the candidate IRS location in cell $i$ (or the BS if $i=0$) can achieve an LoS path with that in cell $i'$ (or the BS if $i'=0$).
To further describe the LoS availability from any candidate IRS location or the BS to all possible user locations within each cell $j, j \in {\cal J}$, we define a virtual node $J+j$ to represent the latter in cell $j$, as shown in Fig. \ref{Fig:systemmodel}(a).
Accordingly, we define additional binary variables $\mu_{i, J+j}$, $\forall i\in {\cal I},\forall j\in {\cal J}$, which is equal to 1 iff the candidate IRS location in cell $i$ (or the BS if $i = 0$) can achieve LoS paths with all possible user locations in cell $j$.
Let ${\cal U} \triangleq \{J, J+1, \ldots, 2J-1\}$ denote the set of all virtual nodes.
It is not difficult to verify that $\mu_{i,i'}=\mu_{i',i}, i, i'\in {\cal I}\cup{\cal U}$.
It should be mentioned that since local LoS coverage can be achieved within each single cell with a BS or candidate IRS location, we have $\mu_{j, J+j} = 1, \forall j\in {\cal I}$.
In practice, the above LoS indicators $\mu_{i, i'}$ can be measured offline in the region interested through various techniques such as ray tracing \cite{Fuschini_2015}.
In this paper, we focus on the downlink from the BS to all possible users in different cells.
Hence, we can set $\mu_{i,0}=0, i \in {\cal I}$ and $\mu_{J+j,i}=0, j \in {\cal J}, i \in {\cal I}\cup{\cal U}$.
The results can be easily extended to the uplink scenario as a direct/cascaded LoS path from the BS to any user in the downlink is also available for its uplink communication to the BS in practice.
\begin{remark}
	It is worth noting that in our previous work \cite{Mei2023IRSdeploy}, since we focused on a network-level performance metric to infer the user communication performance (i.e., the minimum number of PIRS reflections required for the BS to establish an LoS link with each cell for any given PIRS deployment), it suffices to represent the candidate IRS location and all possible user locations in each cell by a single node without accounting for the actual channel conditions. However, this cannot be applied to this work focusing on fine-grained signal-level performance metric (i.e., user SNRs).
\end{remark}

Next, we characterize the LoS channel between any two nodes in the system (if any).
To this end, let $d_{i,i'}$, $i\neq i', i, i' \in {\cal I}$ be the distance between nodes $i$ (BS/IRS) and $i'$ (BS/IRS).
Since there exists an infinite number of possible user locations within each cell $j$, to describe the distances between node $i$ (BS/IRS) and node $J+j$, we consider the worst-case user location in cell $j$ that achieves the largest distance from node $i$.
Accordingly, let $d^{\max}_{i, J+j}, i\in {\cal I}, j \in \mathcal{J}$ denote the distance between node $i$ and its associated worst-case user location in cell $j$ (or node $J+j$).
Define $\bm H_{0, i}, i\in{\cal I}_0$ as the baseband equivalent channel from node 0 (BS) to node $i$ (IRS), $\bm S_{i, i'}, i,i'\in{\cal I}_0, i\neq i'$ as that from node $i$ (IRS) to node $i'$ (IRS), and $\bm g^{\sf H}_{i, J+j}$, $i\in{\cal I}$, $j\in {\cal J}$ as that from node $i$ (BS/IRS) to node $J+j$ (or node $i$'s associated worst-case user location in cell $j$).
Without loss of generality, we assume that the BS and each IRS are equipped with a uniform linear array (ULA) and a uniform planar array (UPA) parallel to the $x$-$z$ plane, respectively.
Moreover, it is assumed that far-field propagation can be achieved over all LoS links; hence, the LoS channel between any two nodes can be modeled as the product of transmit and receive array responses at the two sides.
For the ULA at the BS, its transmit array response with respect to (w.r.t.) IRS $i$ is written as
\begin{eqnarray}
	{\tilde{\bm h}}_{0,i,t} = \bm u(\frac{2d}{\lambda}\cos\varphi^{t}_{0, i}, M )\in \mathbb{C}^{M \times 1},
\end{eqnarray}
where $\varphi^{t}_{0, i}$ denotes the angle-of-departure (AoD) from the BS to IRS $i$, $\lambda$ denotes the signal wavelength, $d$ denotes the spacing between two adjacent antennas/elements at the BS/each IRS, and $\bm u$ is the steering vector function defined as
\begin{eqnarray}
	\bm u (\varsigma, M')= [1, e^{-\jmath\pi\varsigma}, \ldots, e^{-\jmath\pi(M' -1) \varsigma}]^{\sf T} \in \mathbb{C}^{M' \times 1}.
\end{eqnarray}
For the UPA at each IRS, we establish a local coordinate system on it and assume that it is parallel to the $x$-$z$ plane. Hence, transmit/receive array response vector can be expressed as the Kronecker product of two steering vector functions in the $x$- and $z$-axis directions, respectively.
In particular, let $N_{i,x}$ and $N_{i,z}$ denote the number of elements at IRS $i$ in the $x$- and $z$-axis directions, respectively, with $N_i = N_{i,x} \times N_{i,z}$.
Then, the transmit array response of IRS $i$ w.r.t. node $j$ (IRS/user) is expressed as
\begin{align}
	{\tilde{\bm s}}_{i,j,t} =\bm {u}(\frac{2d}{\lambda}\cos(\varphi^{t}_{i, j})\sin(\vartheta^t_{i, j}), N_{i,x}) \!\otimes\! \bm {u}(\frac{2d}{\lambda}\cos(\vartheta^t_{i, j}), N_{i,z}),
\end{align}
where $\varphi^{t}_{i, j}$ and $\vartheta^t_{i, j}$ denote the azimuth and elevation AoDs from IRS $i$ to node $j$, respectively.
Similarly, we can define the receive array response of node $j$ w.r.t. node $i$ (IRS/BS) and denote it as ${\tilde{\bm s}}_{j, i,r}$.

Based on the above, if $\mu_{0,i} = 1$, $i\in {\cal I}_0$, the BS$\to$IRS $i$ LoS channel is expressed as
\begin{eqnarray}\label{Eq:BStoIRS channel}
	\hspace{-1em}	{\bm H}_{0,i} &=& e^{-\frac{\jmath2\pi d_{0,i}}{\lambda}}\kappa_{0,i}{\tilde{\bm s}}_{0,i,r}{\tilde{\bm h}}^{\sf H}_{0,i,t}\in  \mathbb{C}^{N_i \times M},
\end{eqnarray}
where $\kappa_{0,i}\triangleq { \sqrt{\beta_0}}/{d^\frac{\alpha}{2}_{0,i}}$ denotes the LoS path gain from the BS to IRS $i$; $\alpha$ and $\beta_0$ respectively denote the LoS path-loss exponent and path gain at the reference distance of one meter (m).
Similarly, if $\mu_{i, i'} = 1$, $i, i'\in {\cal I}_0$, the IRS $i\to$IRS $i'$ LoS channel is given by
\begin{eqnarray}\label{Eq:IRStoIRS channel}
	{\bm S}_{i,i'} &=& e^{-\frac{\jmath2\pi d_{i,i'}}{\lambda}}\kappa_{i,i'}{\tilde{\bm s}}_{i,i',r}{\tilde{\bm s}}^{\sf H}_{i,i',t}\in  \mathbb{C}^{N_i \times N_{i'}}.
\end{eqnarray}
Finally, if $\mu_{i, J+j} = 1$, $i\in {\cal I}$, $j \in \mathcal{J}$, the LoS channel from node $i$ (BS/IRS) to node $J+j$ is given by
\begin{eqnarray}\label{Eq:IRStoCell channel}
	\bm g^{\sf H}_{i,J+j} =
	\begin{cases}
			\kappa_{0,J+j} e^{-\frac{\jmath2\pi d^{\max}_{0,J+j}}{\lambda}}{\tilde{\bm h}}^{\sf H}_{0,J+j,t}\in  \mathbb{C}^{1 \times M}& i=0,\\
		\kappa_{i,J+j} e^{-\frac{\jmath2\pi d^{\max}_{i,J+j}}{\lambda}}{\tilde{\bm s}}^{\sf H}_{i,J+j,t}\in  \mathbb{C}^{1 \times N_i}&  i\in {\cal I}_0, 
	\end{cases}
\end{eqnarray}
where $\kappa_{i,J+j} \triangleq { \sqrt{\beta_0}}/{(d^{\max}_{i,J+j})^\frac{\alpha}{2}}$ denotes the corresponding worst-case LoS path gain.

Next, we model the considered region and all LoS paths inside it based on a directed LoS graph $G=(V,E)$, where the vertex set $V$ consists of the nodes in ${\cal I}$ and ${\cal U}$, i.e., $V={\cal I} \cup {\cal U}$, and the edge set is given by $E=\{(i,i')|\mu_{i,i'}=1, i \ne i', i, i'\in V\}$, i.e., there is an edge from vertex $i$ to vertex $i'$ iff $\mu_{i,i'}=1$.
Note that for each cell without any candidate IRS location or BS, e.g., cell $i, i \in {\cal J} \backslash {\cal I}$, we only need to consider its possible user locations in $G$, i.e., node $J+i$ with $\{J+i\} \in {\cal U}$.
By this means, we can establish a one-to-one mapping between the LoS path from the BS to any user location in $\cal D$ and a path in $G$.
Fig. \ref{Fig:systemmodel}(c) depicts one graph $G$ generated based on the considered region $\cal D$ in Fig. \ref{Fig:systemmodel}(a) and their pairwise binary LoS indicators $\mu_{i,i'}$'s.
Hereafter, we use vertices and nodes interchangeably and refer to a path as both the LoS path in $\cal D$ and its corresponding path in $G$ without ambiguity.

\section{Deployment Cost and Communication Performance  Characterization}
In this section, we characterize the total IRS deployment cost and the  communication performance in the considered system to reveal the fundamental trade-offs between them.

\subsection{Deployment Cost}
We consider that the total IRS deployment cost depends on both the number of cells selected for IRS deployment and that of tiles deployed in selected cells.
Accordingly, we express the total deployment cost as
\begin{align}\label{Eq:deploy cost}
	c({\cal P},{\cal A}, {\cal T}) =  \underbrace{c_{P, 0} |{\cal P}| + c_{A, 0}|{\cal A}|}_{\text{Cell-use cost}}+ \underbrace{  c_{P}\sum_{p \in {\cal P}}T_{p} + c_{A}\sum_{a\in {\cal A}}T_{a} }_{\text{Hardware cost}},
\end{align}
where $c_{P,0}$ ($c_{A,0}$) denotes a fixed cost for deploying a PIRS (an AIRS) in any cell, and $c_{P}$ ($c_{A}$) denotes the cost per passive (active) tile. 
It is important to note that $c_{P,0}$ ($c_{A,0}$) captures the cell-use cost of deploying a PIRS (AIRS), including mounting cost, controller cost, and power cost etc., which are regardless of the size of the IRS (or number of tiles deployed); while $c_{P}$ ($c_{A}$) captures the hardware cost of deploying a passive (active) tile.
Note that $c_{A} > c_{P}$ and $c_{A, 0}> c_{P, 0}$  generally hold in practice due to the higher power consumption and tile cost (with power amplification) associated with AIRS than PIRS.

\subsection{Communication Performance}

By properly deploying PIRSs/AIRSs in $\cal D$, multiple direct/cascaded LoS links may be achieved from the BS to each cell and the desired user locations within it, which enables us to properly select one or multiple LoS paths to serve them. 
As such, in this paper, we consider the maximum worst-case SNR among all desired user locations within each cell achievable by selecting one LoS path from all possible direct and cascaded LoS paths from the BS to it\footnote{Note that multiple LoS paths can also be selected to serve each user to further enhance its received SNR by exploiting passive beam combining at its receiver \cite{Mei2022MultiPath}. However, to ease the analysis, we only consider a single-path selection for each user, which also provides a tractable performance lower bound on the actual SNR when multiple LoS paths are selected.} as the communication performance metric.
This SNR performance depends on the number and locations of cells deployed with IRSs (i.e., ${\cal P}$ and ${\cal A}$), as well as the number of tiles deployed (i.e., $\cal T$). 
This is because deploying IRSs in more cells may create more signal paths, which helps enhance the coverage of the BS. 
While for any given ${\cal P}$ and ${\cal A}$, by increasing the number of tiles per IRS, the strength of the signal over each multi-IRS-reflection path can be boosted thanks to the improved CPB gain.

Furthermore, for each cell, if a multi-IRS-reflection LoS link needs to be selected for the user locations inside it, we consider the presence of at most one AIRS over it, which ensures a sufficiently high amplification gain of each deployed AIRS and also simplifies the IRS reflection design \cite{Fu2023multi}.
Based on the above, there exist three types of LoS transmissions from the BS to any cell, i.e., direct transmission, hybrid PIRS and AIRS enabled transmission, and all-PIRS enabled transmission, as shown in Fig. \ref{Fig:systemmodel}(b).

Next, we derive the worst-case SNR performance for each cell under each type of transmission over any LoS path for any given ${\cal P}$, ${\cal A}$, and ${\cal T}$.
To this end, we first define $\Omega = \{0, b_1, b_2, \ldots, b_L, J+j\}$ as any LoS path from vertex $0$ (BS) to vertex $J+j, j\in {\cal J}$ (i.e., the worst-case user location in cell $j$ w.r.t. IRS $b_L$) in $G$, where $L\geq 0$ and $b_l \in {\cal P}\cup {\cal A}$ denote the number of intermediate vertices in $\Omega$ and the index of the $l$-th intermediate vertex (PIRS/AIRS).
Since at most one AIRS can exist over any LoS path, it should hold that $|\Omega\cap {\cal A}|\le 1$.
For convenience, we define $b_0 = 0$ and $b_{L+1} = J+j$.
Let $\bm w \in \mathbb{C}^{M\times 1}$ and $P_0$ denote the BS's beamforming vector and transmit power, respectively, with $\|\bm w\|^2 = P_0$.

\subsubsection{Type 1: Direct Transmission} 
In this case, we have $L=0$ and $\mu_{0,J+j} = 1$.
Through the direct link, the received signal at the worst-case user location in cell $j$ (w.r.t. the BS) is given by
\begin{eqnarray}\label{Eq:received signal1}
	y_{J+j} = {\bm g}_{0,J+j}^{\sf H}\bm w s + z,
\end{eqnarray}
where $s$ denotes the transmitted symbol at the BS with $\mathbb{E}[|s|^2] = 1$, and $z\in \mathbb{C}$ denotes the additive white Gaussian noise (AWGN) at the user receiver.
It is assumed that $z\sim \mathcal{CN}(0,\sigma^2)$, where the AWGN power is assumed to be identical to the amplification noise power of each AIRS as $\sigma^2$.
It is known that the maximum-ratio transmission, i.e., $\bm w =\sqrt{P_0}{\tilde{\bm h}_{0, J+j, t}}/{\|\tilde{\bm h}_{0, J+j, t}\|}$, can maximize the received signal power, i.e., $|{\bm g}_{0,J+j}^{\sf H}\bm w|^2$.
Therefore, the worst-case received SNR in cell $j$ in this case is set as
\begin{eqnarray} \label{Eq:SNR-direct}
	{\gamma}_{0, J+j} = C_0\kappa^{2}_{0,J+j},
\end{eqnarray}
where $C_0 = \frac{P_0M}{\sigma^{2}}$. 
It can be observed from \eqref{Eq:SNR-direct} that ${\gamma}_{0, J+j}$ is independent of ${\cal P}$, ${\cal A}$, and ${\cal T}$.
Note that if $\mu_{0,J+j} = 0$, we can set ${\gamma}_{0, J+j} = -\infty$.

\subsubsection{Type 2 : Hybrid PIRS and AIRS Enabled Transmission} 
In this case,  we have $L\geq1$ and $|\Omega \cap {\cal A}| = 1$. Let $b_l$ denote the index of the AIRS node in $\Omega$, i.e., $\Omega  \cap {\cal A} = \{b_l\}$.
Then,  BS$\to$AIRS $b_l$ and the worst-case AIRS $b_l$$\to$cell $j$ multi-reflection channels under $\Omega$ can be respectively expressed as
\begin{eqnarray}
	 \bm h_{0,b_l}(\Omega ) &=& \Big(\prod\nolimits_{m=1}^{l-1}\bm S_{b_m,b_{m+1}}\bm \Phi_{b_m}\Big)\bm H_{0, b_1}\bm w, \label{Eq:Tx-AIRS channel}\\
\tilde{\bm {g}}^{\sf H}_{b_l,J+j}(\Omega) &=& \bm g^{\sf H}_{b_L, J+j}(\prod\nolimits_{m=l}^{L-1}\bm \Phi_{b_{m+1}}\bm S_{b_{m}, b_{m+1}}). \label{Eq:AIRS-Rx channel}
\end{eqnarray}
Through the multi-reflection link $\Omega$, the received signal at the worst-case user location in cell $j$ (w.r.t. IRS $b_L$) is given by
\begin{eqnarray}\label{Eq:received signal}
	\bar{y}_{0, J+j} = \tilde{\bm g}^{\sf H}_{b_l,J+j}(\Omega) \bm \Phi_{b_l } (\bm h_{0,b_l}(\Omega) s + \bm{n}_{b_l}) + z.
\end{eqnarray}	
Then, the corresponding  worst-case received SNR in cell $j$ under $\Omega$ is given by
\begin{eqnarray}\label{Eq:SNR-active}
\bar{\gamma}_{0, J+j}(b_l, \Omega,  {\cal P}, {\cal A}, {\cal T}) 
	= \frac{|\tilde{\bm g}^{\sf H}_{b_l,J+j}(\Omega) \bm \Phi_{b_l} \bm h_{0,b_l}(\Omega) |^2}{|\tilde{\bm g}^{\sf H}_{b_l,J+j}(\Omega)\bm \Phi_{b_l}|^2\sigma^2 +\sigma^2}.
\end{eqnarray}
Let $P_A$ denote the maximum amplification power per reflecting element of the AIRS.
As shown in \cite{Fu2023multi},  to maximize \eqref{Eq:SNR-active}, the optimal AIRS amplification factor, BS beamforming, and each IRS beamforming are respectively given by
\begin{eqnarray}
\hspace{-1em}&&	\eta_{b_l}^2 = \frac{P_{A}}{\kappa^2_{0,b_1}|{\tilde{\bm h}}^{\sf H}_{0,b_1,t}\bm w|^2(\prod_{m=1}^{l-1}\kappa^2_{b_m,b_{m+1}}|A_{b_{m}}|^2)+ \sigma^2}, \label{Eq:AIRS amplitude} \\
\hspace{-1em}&&	\bm w = \sqrt{P_0}{\tilde{\bm h}_{0, b_{1}, t}}/{\|\tilde{\bm h}_{0, b_{1}, t}\|}, \label{Eq:BS beam} \\
\hspace{-1em}&&	\theta_{b_m,n} = \arg([\tilde{\bm s}^{\sf H}_{b_{m-1}, b_m, r}]_n[\tilde{\bm s}_{b_m, b_{m+1}, t}]_n), \forall m, \forall n. \label{Eq:IRS phase}
\end{eqnarray}
It is worth noting that in practice, the beamforming designs in \eqref{Eq:AIRS amplitude}-\eqref{Eq:IRS phase} can be determined by applying distributed and local cooperation among different nodes \cite{Fu2023multi}.

By substituting \eqref{Eq:AIRS amplitude}-\eqref{Eq:IRS phase}  into \eqref{Eq:SNR-active}, \eqref{Eq:SNR-active} is simplified as
\begin{eqnarray}\label{Eq:SNR-active-rewrite}
\hspace{-4em}&&\bar{\gamma}_{0, J+j}(b_l,\Omega, {\cal P},{\cal A},{\cal T}) =    \Big(
\frac{\kappa^{-2}_{0,b_{1}} }{C_0N^2T_{b_l}} \prod_{ m =1}^{l-1}\frac{\kappa^{-2}_{b_{ m},b_{m+1}}}{N^4T^2_{b_{m}}} \nonumber\\
&&\hspace{1em}+  \frac{1}{C_A} \prod_{ m =l}^{L}\frac{\kappa^{-2}_{b_{ m},b_{m+1}}}{N^4T^2_{b_{m}} } + \frac{\kappa^{-2}_{0,b_{1}}}{C_0C_A} \!\prod_{ m =1}^{L}\frac{\kappa^{-2}_{b_{ m},b_{m+1}}}{N^4T^2_{b_{m}} }  \Big) ^{-1}, 
\end{eqnarray}
with $C_A = \frac{P_A}{\sigma^{2}}$. 

For any given $b_l$, ${\cal P}$, ${\cal A}$, and ${\cal T}$, we should select an optimal reflection path $\Omega$ that maximizes \eqref{Eq:SNR-active-rewrite}. 
It has been shown in \cite{Zhang2023MAMP} that such a path selection problem can be equivalently transformed into two subproblems, aiming to select the optimal BS-to-AIRS $b_l$ and AIRS $b_l$-to-cell $j$ sub-paths, respectively. These two subproblems can be further simplified to minimizing $\kappa^{-2}_{0,b_{1}}\prod_{ m =1}^{l-1}\frac{\kappa^{-2}_{b_{m},b_{m+1}}}{N^4T^2_{b_{m}}}$ and $ \prod_{m =l}^{L}\frac{\kappa^{-2}_{b_{m},b_{m+1}}}{N^4T^2_{b_{m}}}$ \cite{Zhang2023MAMP}, respectively, after discarding irrelevant constant scalars.
By further taking their logarithm, it is equivalent to minimizing
\begin{eqnarray} 
	\bar{\lambda}_{0,b_l}(\Omega, {\cal P},{\cal A},{\cal T}) &=&	\ln\kappa^{-2}_{0,b_{1}} \!+\! \sum_{ m =1}^{l-1}	\ln\frac{\kappa^{-2}_{b_{ m},b_{m+1}}}{N^4T^2_{b_{m}} },  \label{Eq:SNR-AIRS-rewrite2}\\
\bar{\lambda}_{b_l,J+j}(\Omega, {\cal P},{\cal A},{\cal T}) &=& \sum_{m =l}^{L}	\ln\frac{\kappa^{-2}_{b_{m},b_{m+1}}}{N^4T^2_{b_{m}} }. \label{Eq:SNR-AIRS-rewrite3}
\end{eqnarray}

Next,  to find the optimal reflection path $\Omega$ that  minimizes both \eqref{Eq:SNR-AIRS-rewrite2} and \eqref{Eq:SNR-AIRS-rewrite3}, we define the following weight function for each edge $(i,i'), (i,i') \in E$,  i.e.,
\begin{eqnarray}\label{Eq:AIRSweight matrix}
	W_{i,i'}(b_l,  {\cal P},{\cal A}, \mathcal{T}) =
 \begin{cases}
		\ln{\kappa^{-2}_{0,i'} } & i = 0, i' \in {\cal I}\cup  {\cal U},\\
			\ln\frac{\kappa^{-2}_{b_l,i'} }{T^2_{b_l}N^4} & i = b_l, i' \in {\cal I}\cup  {\cal U},\\		
		\ln\frac{\kappa^{-2}_{i,i'}}{T^2_{i}N^4 } &i \in {\cal P}, i' \in {\cal I}\cup {\cal U},\\	
		\infty & {\text{otherwise}}.
	\end{cases}
\end{eqnarray}
Given \eqref{Eq:AIRSweight matrix} and the path $\Omega$, it can be verified that 
\begin{eqnarray} 
	\bar{\lambda}_{0,b_l}(\Omega, {\cal P},{\cal A},{\cal T}) &=&	 \sum_{ m =0}^{l-1}	W_{b_m, b_{m+1}}({b_l}, {\cal P},{\cal A}, \mathcal{T}), \label{Eq:pathSumweight1} \\
	\bar{\lambda}_{b_l,J+j}(\Omega, {\cal P},{\cal A},{\cal T}) &=&  \sum_{ m =l}^{L}	W_{b_m, b_{m+1}}({b_l}, {\cal P},{\cal A}, \mathcal{T}), \label{Eq:pathSumweight2} 
\end{eqnarray}
which indicates that the sum of weights over the sub-path from vertex 0 to vertex $b_l$ is equal to \eqref{Eq:SNR-AIRS-rewrite2} and that over the sub-path from vertex $b_l$ to vertex $J+j$ is equal to \eqref{Eq:SNR-AIRS-rewrite3}. It should also be mentioned that we set the weights of several edges to infinity in the fourth case of \eqref{Eq:AIRSweight matrix}. This may happen in the following scenarios. First, if vertex $i$ corresponds to another AIRS (instead of AIRS $b_l$), this will violate the presumption that only a single AIRS is selected over $\Omega$. Second, if vertex $i$ corresponds to the users in cell $i$ or vertex $i'$ corresponds to the BS, then edge $(i, i')$ cannot be used to form any desired LoS path, since the downlink scenario is considered.
Third, if no IRS is deployed in cell $i$, then vertex $i$ should not be selected as an intermediate vertex by any desired LoS path. 
As such, in all the scenarios above, edge $(i, i')$ should be removed. 
However, to keep the topology of $G$, we set their weights to infinity equivalently.           
As a result, if either $\bar{\lambda}_{0,b_l}(\Omega, {\cal P},{\cal A},{\cal T})$ or $\bar{\lambda}_{b_l,J+j}(\Omega, {\cal P},{\cal A},{\cal T}) $ is equal to infinity, 
it implies that one of the two sub-paths involves at least one undesired vertex (e.g., another AIRS or the candidate location without IRS deployed).
Consequently, the path $\Omega$ cannot be used to serve users in cell $j$ in this case.

Let $\Gamma_{0, b_l}$ and $\Gamma_{b_l,J+j}$ denote the sets of all LoS paths from vertex $0$ to vertex $b_l$ and those from vertex $b_l$  to vertex $J+j$ in $G$, respectively.
With the weights defined in \eqref{Eq:AIRSweight matrix}, we select the optimal sub-paths of $\Omega$ to minimize $\bar{\lambda}_{0,b_l}(\Omega, {\cal P},{\cal A},{\cal T})$ and $\bar{\lambda}_{b_l,J+j}(\Omega, {\cal P},{\cal A},{\cal T})$, respectively, i.e.,
\begin{eqnarray}
\hspace{-1.8em}	\bar{\lambda}_{0,b_l}({\cal P},{\cal A},{\cal T}) &=&\min\limits_{\Omega \in \Gamma_{0, b_l}}\bar{\lambda}_{0,b_l}(\Omega, {\cal P},{\cal A},{\cal T}),\label{Eq:AIRS-weight1}\\
\hspace{-1.8em}	\bar{\lambda}_{b_l,J+j}({\cal P},{\cal A},{\cal T}) &=&\min\limits_{\Omega\in \Gamma_{b_l,J+j}}\bar{\lambda}_{b_l,J+j}(\Omega, {\cal P},{\cal A},{\cal T}), \label{Eq:AIRS-weight2}
\end{eqnarray}
both of which can be efficiently calculated on $G$ using classical shortest path algorithms in graph theory (e.g., Bellman-Ford algorithm \cite{west2001introduction}). 
Note that if $\Gamma_{0,b_l} = \emptyset$ or $\Gamma_{b_l,J+j} = \emptyset$, i.e., there is no path from vertex 0  to vertex $J+j$ through vertex $b_l$, we  equivalently set $	\bar{\lambda}_{0,b_l}({\cal P},{\cal A},{\cal T}) = \bar{\lambda}_{b_l,J+j}({\cal P},{\cal A},{\cal T}) =  \infty$.

 By substituting \eqref{Eq:AIRS-weight1} and \eqref{Eq:AIRS-weight2} into \eqref{Eq:SNR-active-rewrite}, we can obtain the maximum worst-case SNR in cell $j$ via hybrid-PIRS/AIRS LoS path selection for any given $b_l$, $\cal P$, $\cal A$, and $\cal T$ as
\begin{eqnarray}\label{Eq:SNR-weight3-AIRS}
\bar{\gamma}_{0, J+j}(b_l, {\cal P},{\cal A},{\cal T})&=& \Big(\frac{e^{\bar{\lambda}_{0,b_l}({\cal P},{\cal A},{\cal T})}}{C_0N^2T_{b_l}}+ \frac{e^{\bar{\lambda}_{b_l,J+j}({\cal P},{\cal A},{\cal T})  }}{C_A}\nonumber\\
&&\hspace{-5em} +	 \frac{ e^{(\bar{\lambda}_{0,b_l}({\cal P},{\cal A},{\cal T}) +\bar{\lambda}_{b_l,J+j}({\cal P},{\cal A},{\cal T}) )}}{C_0C_A}\Big)^{-1}, j \in {\cal J}.
\end{eqnarray}
Then, the maximum worst-case SNR in cell $j$ via hybrid-PIRS/AIRS LoS path selection for given ${\cal P}$, ${\cal A}$, and ${\cal T}$ can be obtained by comparing \eqref{Eq:SNR-weight3-AIRS} under different $b_l, b_l \in {\cal A}$, i.e.,
\begin{eqnarray}\label{Eq:SNR-weight-AIRS}
	\bar{\gamma}_{J+j}({\cal P},{\cal A},{\cal T})&=&\max\limits_{b_l \in {\cal A}}\bar{\gamma}_{0, J+j}(b_l, {\cal P},{\cal A},{\cal T}), j \in {\cal J}.
\end{eqnarray}

\subsubsection{Type 3: All-PIRS Enabled Transmission}
In this case, we have $L\geq 1$ and $\Omega \cap {\cal A} = \emptyset$. Consequently, the worst-case BS$\to$cell $j$ multi-reflection channel is expressed as
\begin{align}
	&&	\hspace{-1.5em}\tilde{ {g}}_{0,J+j}(\Omega) = \bm g^{\sf H}_{b_L, J+j}\bm \Phi_{b_L} \Big(\prod\nolimits_{m=1}^{L-1}\bm S_{b_m,b_{m+1}}\bm \Phi_{b_k}\Big)\bm H_{0, b_1}\bm w.
	\label{Eq:Tx-Rx channel}
\end{align}
To maximize the end-to-end channel power gain $|\tilde{ {g}}_{0,J+j}(\Omega)|^2$, it can be shown that the optimal beamforming at the BS and IRSs should satisfy \eqref{Eq:BS beam} and \eqref{Eq:IRS phase}, respectively.
Therefore, in this case, the worst-case received SNR in cell $j$ under $\Omega$ is given by
\begin{eqnarray} \label{Eq:SNR-all-rewrite}
	\tilde{\gamma}_{0, J+j}(\Omega, {\cal P},{\cal A},{\cal T}) = \left( \frac{\kappa^{-2}_{0,b_1}}{C_0 } \prod_{ m =1}^{L}\frac{\kappa^{-2}_{b_{ m},b_{m+1}}}{N^4T^2_{b_{m}} } \right) ^{-1}.
\end{eqnarray}
It has been shown in \cite{Mei2021routing} and \cite{Mei2022MIMO} that selecting the optimal path for maximizing \eqref{Eq:SNR-all-rewrite} is equivalent to minimizing
\begin{eqnarray} \label{Eq:SNR-all-rewrite2}
	\tilde{\lambda}_{0,J+j}(\Omega, {\cal P},{\cal A}, {\cal T}) = \ln\kappa^{-2}_{0,b_{1}} + \sum_{ m =1}^{L}	\ln\frac{\kappa^{-2}_{b_{m},b_{m+1}}}{N^4T^2_{b_{m}} }.
\end{eqnarray}

Next, we show that minimizing \eqref{Eq:SNR-all-rewrite2} can still be equivalently reformulated as a shortest path problem under proper weight assignment, similar to \eqref{Eq:AIRSweight matrix}.
However, different from \eqref{Eq:AIRSweight matrix}, the desired path $\Omega$ in $G$ must not involve any AIRS nodes.
Accordingly, for each $(i,i')\in E$, we modify the weights in \eqref{Eq:AIRSweight matrix} as,
\begin{eqnarray}\label{Eq:weight matrix2}
	\tilde{W}_{i,i'}({\cal P},{\cal A}, \mathcal{T}) =
	 \begin{cases}
		\ln{\kappa^{-2}_{0,i'} } &i = 0, i' \in {\cal I}, \\
		\ln\frac{\kappa^{-2}_{i,i'}}{T^2_{i}N^4 } &i \in {\cal P},i' \in {\cal I}\cup {\cal U}, \\
		\infty &{\text{otherwise}}.
	\end{cases}
\end{eqnarray}
Given \eqref{Eq:weight matrix2} and the path $\Omega$, it can be verified that  
\begin{eqnarray} 
	\tilde{\lambda}_{0,J+j}(\Omega, {\cal P},{\cal A},{\cal T}) &=&	 \sum_{ m =0}^{L}	\tilde{W}_{b_m, b_{m+1}}({\cal P},{\cal A}, \mathcal{T}).
\end{eqnarray}
Note that in addition to the fourth case in \eqref{Eq:AIRSweight matrix}, we also set the weight of an edge to infinity in \eqref{Eq:weight matrix2} in the following scenarios.  
First, we set the weights of all edges starting from a vertex $i, i\in {\cal A}$ to infinity, as any AIRS node cannot be involved in $\Omega$.
Second, if $(0, J+j)\in E $ (or $\mu_{0,J+j} = 1$),
we also set its weight  to infinity, as this corresponds to the direct LoS transmission from the BS to the users in cell $j$, which has already been accounted for in Section III-B1.
Similarly to \eqref{Eq:pathSumweight1} and \eqref{Eq:pathSumweight2}, if $\tilde \lambda_{0,J+j}(\Omega, {\cal P},{\cal A},{\cal T})$ is equal to infinity, it implies that the path $\Omega$ cannot be used to serve users in cell $j$.

Let $\Gamma_{0,J+j}$ denote the set of all LoS paths from vertex $0$ to vertex $J+j, j \in {\cal J}$ in $G$.
With the weights defined in \eqref{Eq:weight matrix2}, we select the optimal $\Omega$ to minimize $\tilde{\lambda}_{0,J+j}(\Omega, {\cal P},{\cal A},{\cal T}) $, i.e.,
\begin{eqnarray}\label{Eq:weight1}
	\tilde{\lambda}_{0,J+j}({\cal P},{\cal A},{\cal T})= \min\limits_{\Omega \in \Gamma_{0,J+j}}\tilde{\lambda}_{0,J+j}(\Omega, {\cal P},{\cal A},{\cal T}),
\end{eqnarray}
which can still be calculated using the shortest path algorithms in graph theory.
Note that if $\Gamma_{0,J+j} = \emptyset$, we  can equivalently set $\tilde{\lambda}_{0,J+j}({\cal P},{\cal A},{\cal T}) =  \infty$.

By substituting \eqref{Eq:weight1} into \eqref{Eq:SNR-all-rewrite}, we can obtain the maximum worst-case SNR in cell $j$ via the all-PIRS-reflection path selection for any given $\cal P$, $\cal A$, and $\cal T$ as
\begin{eqnarray}\label{Eq:SNR-weight-PIRS}
	\tilde{\gamma}_{J+j}({\cal P},{\cal A},{\cal T}) =e^{-	\tilde\lambda_{0,J+j}({\cal P},{\cal A},{\cal T})}C_0, j \in {\cal J}.
\end{eqnarray}

Finally, by comparing the maximum worst-case SNRs under the above three types of transmissions in \eqref{Eq:SNR-direct}, \eqref{Eq:SNR-weight-AIRS}, and \eqref{Eq:SNR-weight-PIRS}, respectively, we set the maximum worst-case SNR for cell $j$ under any  given ${\cal P}$, ${\cal A}$, and ${\cal T}$ as
\begin{eqnarray}\label{Eq:maxSNR}
  \hspace{-2em}	\gamma_{J+j}({\cal P},{\cal A},{\cal T}) &=& \nonumber \\
	&& \hspace{-8em}\max\{\gamma_{0,J+j}, \bar{\gamma}_{J+j}({\cal P},{\cal A},{\cal T}), \tilde{\gamma}_{J+j}({\cal P},{\cal A},{\cal T})\}, j \in {\cal J}.
\end{eqnarray}
It is noted that if $\gamma_{J+j}({\cal P},{\cal A},{\cal T}) > 0, \forall j \in {\cal J}$, each user in cell $j$ can communicate with the BS via an LoS link (either direct, all PIRSs, or hybrid PIRSs and AIRSs).
Otherwise, it is infeasible to achieve global LoS coverage in ${\cal D}$ with the given ${\cal A}$, ${\cal P}$, and ${\cal T}$.

\subsection{Cost-Performance Trade-offs}

In this subsection, we show that there exist two trade-offs between maximizing the SNR in \eqref{Eq:maxSNR} and minimizing the total deployment cost in \eqref{Eq:deploy cost} (i.e.,  $c({\cal P},{\cal A}, {\cal T})$).

First, for any given ${\cal P}$ and ${\cal A}$, the cell-use cost in \eqref{Eq:deploy cost} should be fixed.
If we deploy more tiles per IRS, i.e.,  increase $T_i$'s, the hardware cost in \eqref{Eq:deploy cost} will increase, resulting in a higher total deployment cost.
On the other hand, this also helps improve the worst-case SNRs in \eqref{Eq:SNR-active-rewrite} and  \eqref{Eq:SNR-all-rewrite} (and thus \eqref{Eq:maxSNR}) thanks to the more significant CPB gain that can be reaped from each multi-reflection path.
It thus follows that {\it the tile number for each PIRS/AIRS should optimally balance between the hardware cost in \eqref{Eq:deploy cost} and SNR performance.}

Second, for any given total numbers of passive and active tiles (i.e., $\sum_{p\in {\cal P}} T_p$ and $\sum_{a\in {\cal A}} T_a$ ), the hardware cost in \eqref{Eq:deploy cost} should be fixed. 
If we distribute these tiles in more cells, i.e., increase $|\cal P|$ and/or $|\cal A|$, the cell-use cost in \eqref{Eq:deploy cost} will increase, resulting in a higher total deployment cost. 
Nonetheless, by this means, the IRSs can cover a larger portion of $\cal D$, which helps create a larger number of cascaded LoS paths. 
This thus results in a larger size of  $\Gamma_{0, b_l}$ and $\Gamma_{b_l, J+j}$ in \eqref{Eq:AIRS-weight1} and \eqref{Eq:AIRS-weight2}, as well as $\Gamma_{0,J+j}$ in \eqref{Eq:weight1}, which is beneficial to improve the SNR performance in \eqref{Eq:maxSNR} thanks to the enhanced LoS path diversity. 
Furthermore, even for a fixed number of cells used for IRS deployment (i.e., $|\cal P| + |\cal A|$), the cell-use cost (and the total deployment cost) in \eqref{Eq:deploy cost} may also be increased by increasing $|\cal A|$ due to $c_{A,0} > c_{P,0}$.
This, on the other hand, may also help improve the SNR performance in \eqref{Eq:maxSNR}, thanks to the AIRS's capability of amplifying the incident signal \cite{Fu2023multi}. 
It thus follows that {\it the number of cells used for PIRS/AIRS deployment should optimally balance between the cell-use cost in \eqref{Eq:deploy cost} and SNR performance.}

To validate this trade-off, we next consider a simplified example, where the BS transmits signal to cell $j$ over a multi-IRS-reflection path formed by $L$ IRSs, denoted as $\hat{\Omega} = \{0, b_1, \ldots, b_l, \ldots, b_L, J+j\}$.
We assume that the distance between any two adjacent nodes in $\hat{\Omega}$ is identical to $d_0$ and consider the following two cases. 
In the first case, all IRSs over $\hat{\Omega}$ are PIRSs (each equipped with $T_0$ tiles).
While for the second case, PIRS $b_l$ in the first case is replaced by an AIRS equipped with $T_0c_{P,0}/{c_{A,0}}$ (assumed to be an integer for convenience) tiles, such that its total hardware cost is the same as the first case, while its cell-use cost increases. 
Next, we show that the second case may achieve a higher received SNR in cell $j$ under $\hat\Omega$ than the first case.
Specifically, in the first case, it follows from \eqref{Eq:SNR-all-rewrite} that the received SNR in cell $j$ under $\hat{\Omega}$ is given by
\begin{eqnarray} \label{Eq:SNR-all-rewrite1}
	{\gamma}_{P}  = C_0 \kappa^{2(L+1)}_{0}{(N^2T_{0})^{2L} },
\end{eqnarray}
where $\kappa_0\triangleq\sqrt{\beta_0}/{d_{0}^\frac{\alpha}{2}}$ denotes the LoS path gain between any two adjacent nodes over $\hat\Omega$.
In the second case, it follows from \eqref{Eq:SNR-active-rewrite} that the received SNR in cell $j$ under $\hat{\Omega}$ is given by
\begingroup\makeatletter\def\f@size{9}\check@mathfonts\def\maketag@@@#1{\hbox{\m@th\normalsize\normalfont#1}}
\begin{align}
	\!\!\!\!\!	{\gamma}_{A} \!=\!	\Big(
	\frac{\kappa^{-2l}_{0}}{C_0c'N_0N_0^{2(l-1)}} 
	\!+\!   \frac{\kappa_0^{-2(L-l+1)}}{C_Ac'^2N_0^{2(L-l+1)}}  \!+ \!\frac{\kappa_0^{-2(L+1)}}{C_0C_Ac'^2N_0^{2L}}  \Big) ^{-1},
\end{align}\normalsize
where $c' = c_{P,0}/{c_{A,0}}$ and $N_0 = N^2T_{0}$. By comparing ${\gamma}_{P}$ and ${\gamma}_{A}$, we have
\begin{align}\label{Eq:T_p}
	\frac{{\gamma}_{A}}{{\gamma}_{P}} = \frac{C_Ac'^2N_0^2}{C_Ac'N_0(\kappa_0N_0)^{2(L+1-l)} + C_0(\kappa_0N_0)^{2l}+N^2_0}.
\end{align}
It is observed that the right-hand side (RHS) of \eqref{Eq:T_p}  decreases with $N_0$ (or $T_0$)  and increases with $\kappa^{-1}_0$ (or $d_0$), which implies that the second case can achieve a higher received SNR than the first case if $T_0$ is small and/or $d_0$ is large, thus validating our previous claim.


\section{Problem Formulation }
To resolve the fundamental trade-offs shown in Section III-C, in this paper, we aim to jointly optimize the PIRS and AIRS deployment (i.e., ${\cal A}$ and ${\cal P}$) and the deployed tile number per candidate location (i.e., ${\cal T}$) to minimize the total deployment cost in \eqref{Eq:deploy cost}, subject to constraints on the SNR performance in \eqref{Eq:maxSNR}. The associated optimization problem is formulated as
\begin{eqnarray} \label{Problem1}
	\text{(P1)}~ \mathop{\text{minimize}}\limits_{{\cal P},{\cal A},\mathcal{T}} && c({\cal P},{\cal A}, \mathcal{T}) \nonumber\\
	\text{subject to}&& {\cal A} \subseteq {\cal I}_0, {\cal P} \subseteq {\cal I}_0, {\cal P} \cap {\cal A} = \emptyset, \label{Cons: Location2}\\
	&&\gamma_{J+j}({\cal P},{\cal A},{\cal T}) \ge \gamma_0, \forall j \in {\cal J}, \label{Cons: SNR} \\
	&& 0< T_{p} \leq T_{0}^{\max}, T_{p} \in \mathbb{N}^+, \forall p \in {\cal P}, \label{Cons: PIRS Tile} \\
	&& 0< T_{a} \leq T_{0}^{\max}, T_{a} \in \mathbb{N}^+, \forall a \in {\cal A}, \label{Cons: AIRS Tile}
\end{eqnarray}
where $\gamma_0 \ge 0$ denotes a prescribed SNR target.

Note that (P1) is a non-convex combinatorial optimization problem with discrete variables and non-convex constraints in \eqref{Cons: SNR}, which is challenging to solve optimally using conventional optimization algorithms in general.
A straightforward approach to optimally solve (P1) is by enumerating all possible PIRS and AIRS location combinations as well as their tile number combinations, which, however, results in practically formidable computational complexity.
In Section \ref{Sec:Solution}, we propose an efficient partial enumeration method that properly discards some undesirable solution sets that cannot achieve optimal performance, thereby significantly reducing the overall computational complexity.

Note that for (P1) to be feasible, a maximum $\gamma_0$ should exist due to the maximum tile number constraints in \eqref{Cons: PIRS Tile} and \eqref{Cons: AIRS Tile}.
In the special case of (P1) where only PIRSs are allowed to be deployed with ${\cal A} = \emptyset$, the maximum SNR in \eqref{Eq:maxSNR} or the maximum SNR target $\gamma_{0}$ in \eqref{Cons: SNR} can be achieved by deploying PIRSs in all cells in ${\cal I}_0$ with each equipped with the maximum number of (passive) tiles, i.e., ${\cal P} = {\cal I}_0$ and $T_i = T_0^{\max}, i \in {\cal P}$.
In the general case with joint PIRS and AIRS deployment, it is expected that a larger $\gamma_0$ should be allowed thanks to the power amplification of AIRS.

\section{Proposed Solution to (P1)}\label{Sec:Solution}
In this section, we first optimize the tile numbers for any given IRS locations, based on which a partial enumeration algorithm is proposed to solve (P1).

\vspace{-1em}
\subsection{Tile Number Optimization}
First, we consider the tile number optimization for any given locations of PIRSs and AIRSs (i.e., ${\cal P}$ and ${\cal A}$). In this case, (P1) can be simplified as 
\begin{eqnarray}
	\text{(P2)} ~	\mathop{\text{minimize}}\limits_{{\cal T}} && c_{P}\sum_{p \in {\cal P}}T_{p} + c_{A}\sum_{a\in {\cal A}}T_{a} +C \nonumber\\
		\text{subject to}
	&&\gamma_{J+j}({\cal P},{\cal A},{\cal T}) \ge \gamma_0,   \forall j \in {\cal J},           \label{Cons: SNR2} \\
	&&  \eqref{Cons: PIRS Tile} \text{~and~} \eqref{Cons: AIRS Tile}, \nonumber
\end{eqnarray}
where $C \triangleq c_{P, 0} |{\cal P}| + c_{A, 0}|{\cal A}|$ denotes the cell-use cost that is a constant.
Since the maximum SNR can be achieved by deploying the maximum number of tiles at each IRS location, (P2) is feasible iff $\gamma_{J+j}({\cal P},{\cal A},{\cal T}^{\max}) \geq  \gamma_0$, $\forall j \in {\cal J}$, where 
  ${\cal T}^{\max} = \{T_{i}| T_i = T_0^{\max}, i\in {\cal P}\cup {\cal A}\}$.

However, (P2) is still a non-convex combinatorial optimization problem.
Although it can be optimally solved by enumerating all possible tile number combinations, this results in the worst-case complexity in the order of $(T_{0}^{\max})^{|{\cal P}|+ |{\cal A}|}$, which becomes unaffordable if $T_{0}^{\max}$ and/or $|{\cal P}|+ |{\cal A}|$ is practically large. To avoid such high computational complexity, we propose a sequential refinement algorithm to solve (P2), which first obtains an approximate solution by relaxing it into a convex optimization problem. 
Then, we sequentially refine the number of tiles deployed in each cell based on this approximate solution, as detailed below.

\subsubsection{Relaxing (P2) into a Convex Optimization Problem}
The main challenge in solving (P2) arises from the non-convex constraints in \eqref{Cons: SNR2}.
Particularly, such constraints cannot be transformed into a convex form by simply relaxing $T_i$ to be continuous variable, due to the ``max" and ``min" operators involved in $\gamma_{J+j}({\cal P},{\cal A},{\cal T})$ arising from path selection in both Type 2 and Type 3 transmissions (see Section III-B).
To tackle this challenge, we consider simplifying $\gamma_{J+j}({\cal P},{\cal A},{\cal T})$ by  considering only a single path from the BS to each cell in each of Type 2 and Type 3 transmissions, so as to avoid path selection and thus the ``max'' and ``min'' operations involved.
To determine such a single path, we propose to set the tile numbers at all IRS locations as their minimum (i.e., ${\cal T}$ = ${\cal T}^{\min} \triangleq \{T_{i}| T_{i} = 1,  i\in {\cal P}\cup {\cal A} \}$) and then perform the path selection procedures in \eqref{Eq:AIRS-weight1}-\eqref{Eq:SNR-weight-AIRS} and \eqref{Eq:weight1} for Type 2 and Type 3 transmissions, respectively.
Such a single path selection is motivated by the fact that the overall SNR performance in \eqref{Eq:maxSNR} monotonically increases with $T_i$ due to the increasing CPB and amplification gains by PIRSs and AIRSs, respectively.
The effect of these reflection gains can therefore be eliminated by setting their tile number to the minimum, which enables us to find paths with sufficiently high end-to-end LoS path gains and low noise power amplification.
Although this may reduce the size of the feasibility solution set of (P2), it greatly simplifies (P2), and we will also further refine its resultant performance sequentially in the next part of this subsection.

After setting ${\cal T}$ to ${\cal T}^{\min}$,
based on \eqref{Eq:SNR-weight-AIRS} and \eqref{Eq:SNR-weight-PIRS},  we obtain the resulting maximum worst-case SNRs in cell $j$ with Type 2 and Type 3 transmissions as $\bar{\gamma}_{J+j}({\cal P},{\cal A},{\cal T}^{\min})$ and $\tilde{\gamma}_{J+j}({\cal P}, {\cal A},{\cal T}^{\min})$, respectively, and denote their selected paths as $\bar\Omega_j$ and $\tilde\Omega_j$.
Then, based on \eqref{Eq:SNR-active-rewrite} and \eqref{Eq:SNR-all-rewrite}, it can be shown that in actual communications, the worst-case received SNRs over the two paths $\bar\Omega_j$ and $\tilde\Omega_j$ can be respectively expressed as
\begin{eqnarray}
	&&	\hspace{-2em} \bar{\gamma}_{0, J+j}(b_l,\bar\Omega_j, {\cal P},{\cal A},{\cal T}) =\nonumber \\
	&& \Big(
	\frac{\bar{C}_0}{N^2T_{b_l}}\! \prod_{ m =1}^{l-1}\!\!\frac{1}{T^2_{b_{m}}}
	\!+\! {\bar{C}_A}\!\prod_{ m =l}^{L}\!\!\frac{1}{T^2_{b_{m}} } \!+ \!{\bar{C}_0\bar{C}_A} \!\prod_{ m =1}^{L}\!\!\frac{1}{T^2_{b_{m}} } \Big) ^{-1}, \label{Eq:SNR-active-rewrite1}\\
	&& \hspace{-2em} \tilde{\gamma}_{0, J+j}(\tilde \Omega_j, {\cal P},{\cal A},{\cal T}) =  {\tilde{C}_0}^{-1}\prod_{ m =1}^{L}{T^2_{b_{m}} } , \label{Eq:SNR-passive-rewrite1}
\end{eqnarray}
where we assume that $\bar{\Omega}_j\cap {\cal A} = \{b_l\}$ in \eqref{Eq:SNR-active-rewrite1}, $\bar{C}_0 = \frac{e^{\bar{\lambda}_{0,b_l}(\bar{\Omega}_j, {\cal P},{\cal A},{\cal T}^{\min})}}{C_0}$, $\bar{C}_A = \frac{e^{\bar{\lambda}_{b_l,J+j}(\bar{\Omega}_j, {\cal P},{\cal A},{\cal T}^{\min})}}{C_A}$, and $\tilde{C}_0 = \frac{e^{\tilde{\lambda}_{0,J+j}(\tilde{\Omega}_j, {\cal P},{\cal A},{\cal T}^{\min})}}{C_0}$.
Furthermore, in the case of Type 1 transmission, i.e., there is a direct LoS path from the BS to cell $j$, i.e., $\{0,J+j\}$, the actual worst-case received SNR in cell $j$ is given by $\gamma_{0,J+j}$ in \eqref{Eq:SNR-direct}.

Next, we replace $\gamma_{J+j}({\cal P}, {\cal A},{\cal T})$  in \eqref{Cons: SNR2} with the actual worst-case received SNR in cell $j$ over the following path,
\begin{eqnarray}\label{Eq:path}
	\Omega_j =
	\begin{cases}
		{\bar \Omega_j}	& \text{if~} \bar{\gamma}_{J+j}({\cal P},{\cal A},{\cal T}^{\min}) \geq \\
		& \max \{\gamma_{0, J+j}, \tilde{\gamma}_{J+j}({\cal P},{\cal A},{\cal T}^{\min}) \},\\
		\tilde \Omega_j & \text{if~}   \tilde{\gamma}_{J+j}({\cal P},{\cal A},{\cal T}^{\min})>  \\
		&~~\max \{\gamma_{0, J+j}, \bar{\gamma}_{J+j}({\cal P},{\cal A},{\cal T}^{\min})\},  \\		
		\{0,J+j\}	 & {\text{otherwise}},
	\end{cases}
\end{eqnarray}
i.e., the path corresponding to the maximum worst-case received SNR under ${\cal T}= {\cal T}^{\min}$ among the three types of LoS transmission.
Hence, $\gamma_{J+j}({\cal P}, {\cal A},{\cal T})$  in \eqref{Cons: SNR2} becomes
\begin{eqnarray}\label{Eq:SNR-giventile}
\!\!\!\!	\hat \gamma_{J+j}( {\cal P}, {\cal A}, {\cal T}) =
	\begin{cases}
	\bar{\gamma}_{0, J+j}(b_l,\bar\Omega_j, {\cal P},{\cal A},{\cal T}) & \!\!\text{if~}   \Omega_j = \bar{\Omega}_j,  \\		
		\tilde{\gamma}_{0, J+j}(\tilde \Omega_j, {\cal P},{\cal A},{\cal T}) & \!\!\text{if~} \Omega_j = \tilde{\Omega}_j, \\
		\gamma_{0, J+j} & \!\!\text{otherwise},
	\end{cases}
\end{eqnarray}
and the simplified version of (P2) is given by
\begin{eqnarray}
	\text{(P2.1)}~ \mathop{\text{minimize}}\limits_{{\cal T}} && c_{P}\sum_{p \in {\cal P}}T_{p} + c_{A}\sum_{a\in {\cal A}}T_{a} +C \nonumber\\
	\text{subject to}
	&&\hat{\gamma}_{J+j}({\cal P},{\cal A},{\cal T}) \ge \gamma_0,   \forall j \in {\cal J},           \label{Cons: SNR3} \\
	&&  \eqref{Cons: PIRS Tile} \text{~and~} \eqref{Cons: AIRS Tile}. \nonumber
\end{eqnarray}

Although (P2.1) is still a non-convex discrete problem, it can be efficiently relaxed into a convex one. To this end, we define a set of variables $x_{i} \triangleq \ln(T_{i}), i \in {\cal P}\cup {\cal A}$.
By substituting them into \eqref{Eq:SNR-active-rewrite1} and \eqref{Eq:SNR-passive-rewrite1}, $\hat\gamma_{J+j}( {\cal P},{\cal A}, {\cal T})$ in \eqref{Eq:SNR-giventile} can be rewritten as 
\begin{align}\label{Eq:SNR-giventile2}
	\hat \gamma_{J+j}( {\cal P},{\cal A}, \bm x) \!=\!
	\begin{cases}
		\Big(\frac{\bar{C}_0}{N^2}e^{-x_{b_l}-\sum_{m=1}^{l-1}2x_{b_m}}+ \bar{C}_Ae^{-\sum_{m=l}^{L}2x_{b_m}} &~	\\
		~~+\bar{C}_0\bar{C}_Ae^{-\sum_{m=1}^{L}2x_{b_m}}\Big)^{-1} &\hspace{-6em}\text{if~}   \Omega_j = \bar{\Omega}_j,  \\		
	\tilde{C}^{-1}_0e^{\sum_{m=1}^L2x_{b_m}}& \hspace{-6em}\text{if~} \Omega_j = \tilde{\Omega}_j, \\
		\gamma_{0, J+j} &\hspace{-6em}\text{otherwise}.\\
	\end{cases}
\end{align}
By substituting \eqref{Eq:SNR-giventile2} into (P2.1) and relaxing each $x_i$ into a continuous variable, we obtain the following optimization problem w.r.t. $\bm x$, i.e.,
\begin{eqnarray}
	\text{(P2.2)}~ \mathop{\text{minimize}}\limits_{\bm x} &&  c_{P}\sum_{p \in {\cal P}}e^{x_{p}} + c_{A}\sum_{a \in {\cal A}}e^{x_{a}} + C   \nonumber\\
	\text{subject to}
	&&\hat{\gamma}_{J+j}({\cal P},{\cal A},\bm x)  \ge \gamma_0,   \forall j \in {\cal J},          \label{Cons: SNR-re} \\
	&& 0 \leq x_{p} \leq \ln T_{0}^{\max},  \forall p \in {\cal P},  \label{Cons: PIRS Tile-re} \\
	&& 0 \leq x_{a} \leq \ln T_{0}^{\max},  \forall a\in {\cal A}. \label{Cons: AIRS Tile-re} 
\end{eqnarray}
To solve (P2.2), it is noted that \eqref{Cons: SNR-re} can be equivalently transformed into $\hat{\gamma}^{-1}_{J+j}({\cal P},{\cal A},\bm x)  \le \gamma^{-1}_0$, with $\hat{\gamma}_{J+j}^{-1}({\cal P},{\cal A},\bm x)$ being convex w.r.t $\bm x$ based on \eqref{Eq:SNR-giventile2}. As a result, (P2.2) is a convex problem, which can be optimally solved by the interior-point algorithm \cite{boyd2004convex}.

Let $\bm x^\ast\triangleq \{x_i^\ast\}$ denote the optimal solution to (P2.2). Next, we reconstruct an integer $T_i$ from $x^\ast_i$ as
\begin{eqnarray}\label{Eq:1}
{\cal T^\ast} = \{T_{i}^\ast	|T_i^\ast = \lceil e^{x^\ast_{i}} \rceil, i\in {\cal P}\cup {\cal A} \},
\end{eqnarray}
where $\lceil x \rceil$ denotes the least integer greater than or equal to $x$.
Note that ${\cal T^\ast}$ must be a feasible solution to (P2.1) since  $\hat\gamma_{J+j}( {\cal P},{\cal A}, {\cal T}) $ in \eqref{Eq:SNR-giventile} is an element-wise increasing function of $\cal T$, and $T_i^\ast \ge e^{x_i^\ast}, \forall i \in {\cal P} \cup {\cal A}$ holds.
Furthermore, it can be easily shown that ${\cal T^\ast}$ is also a  feasible solution to (P2) by noting $\gamma_{J+j}({\cal P},{\cal A},{\cal T})\geq \hat\gamma_{J+j}({\cal P},{\cal A},{\cal T})$ due to the single path selection in \eqref{Eq:path} without accounting for all available paths. 
In this regard,  $ c({\cal P},{\cal A}, \cal T^\ast)$ can serve as an upper bound on the optimal value of (P2). 
\begin{algorithm}[t]
	\caption{Sequential Refinement Algorithm for (P2)}\label{Algo:tile}
	\begin{algorithmic}[1]
		\STATE Input: ${\cal P}$, ${\cal A}$, and $\gamma_{0}$.
		\STATE Initialize: ${\cal T}^{\max} = \{T_{i}| T_i = T_0^{\max}, i\in {\cal P}\cup {\cal A}\}$.
		\IF {$\gamma_{J+j}({\cal P},{\cal A},{\cal T}^{\max}) \geq   \gamma_{0}, \forall j \in {\cal J}$}
		\STATE  Calculate ${\cal T}^\ast$ in \eqref{Eq:1} via solving (P2.2). 
		\STATE Calculate $\delta_i^\ast = \lceil e^{x^\ast_{i}} \rceil - e^{x^\ast_{i}}, i \in {\cal P} \cup {\cal A}$ and set $s =\arg \max_{i\in {\cal P} \cup {\cal A}} \delta^\ast_i$.
		\STATE Initialize: ${\cal I'} = \emptyset$.
		\WHILE{$|{\cal I'}| \leq |{\cal P}|+|{\cal A}|$}		
		\STATE  Let $t=1$.
		\WHILE{$t \le T^\ast_{s}-1$}
		\STATE  Calculate ${\cal T}_t'$ in \eqref{Eq:2} via solving (P2.4.$t$). 
		\IF {$c({\cal P},{\cal A}, {\cal T}_t') < c({\cal P},{\cal A}, \cal T^\ast)$}
		\STATE Update  ${\cal T}^\ast  = {\cal T}_t'$ and $\delta_i^\ast = \lceil e^{x'_{t,i}} \rceil - e^{x'_{t,i}},  i \in {\cal P} \cup {\cal A}$.
		\STATE Update $t = t+1$.
		\ELSE
		\STATE Break.
		\ENDIF		
		\ENDWHILE
		\STATE Update ${\cal I'} = {\cal I'}\cup \{s\}$ and  $s=\arg \max_{i\in {\cal P} \cup {\cal A}\backslash {\cal I'}} \delta^\ast_i$.
		\ENDWHILE		
		\ELSE
		\STATE (P2) is infeasible.
		\ENDIF		
		\STATE Output ${\cal T}^\ast$ as the optimized solution to (P2).
	\end{algorithmic}
\end{algorithm}
\subsubsection{Refining the Reconstructed Solution}
Although ${\cal T^\ast}$ is a feasible solution to (P2), it might not be optimal due to the round-up operation in \eqref{Eq:1} and the single-path selection in \eqref{Eq:path}.
Next,  to further improve the performance, we propose to sequentially refine the number of tiles deployed in each cell based on ${\cal T^\ast}$.
Specifically, we denote by ${\cal I'} , {\cal I'} \in {\cal P} \cup {\cal A}$ the set of cells whose deployed tile numbers have been refined after the current iteration and need to refine the tile number in cell $s$, $s \in {\cal P} \cup {\cal A}  \backslash {\cal I'}$ in the next iteration. 
The procedures of how to determine $s$ will be specified later.
To this end, we successively remove one tile from its $T^\ast_s$ tiles and at the same time optimize each of $T_i, i \in {\cal P} \cup {\cal A} \backslash {\cal I}', i \ne s$ to minimize the total deployment cost $ c({\cal P},{\cal A}, \cal T)$, with each of $ T_i, i \in {\cal I'}$ being fixed as $T^\ast_i$, until $ c({\cal P},{\cal A}, \cal T)$ cannot be further reduced.
Note that this successive tile removal is different from that in our previous work \cite{Mei2023IRSdeploy}, where each $T_i, i \in {\cal P} \cup {\cal A}, i \ne s$ is also fixed as $T_i^\ast$ (instead of being optimized), since this may always result in an infeasible problem due to the SNR constraints in (P2).
In particular, when the $t$-th tile is removed in cell $s$, i.e., $T_{s}=T_{s}^\ast -t, 1 \le t \le T_{s}^\ast-1$, the associated optimization problem is given by
\begin{eqnarray}
\!\!\!\!\!\!\!\!	\text{(P2.3.$t$)} ~\mathop{\text{minimize}}\limits_{\cal T} && c_{P}\sum_{p \in {\cal P}}T_{p} + c_{A}\sum_{a \in {\cal A}}T_{a} + C \nonumber\\
	\text{subject to}
	&&\hat{\gamma}_{J+j}( {\cal P},{\cal A},{\cal T})  \ge \gamma_0,  \forall j\in {\cal J},            \\
	&& 0 < T_{i} \leq  T_{0}^{\max},  \forall i \in {\cal P} \cup {\cal A}\backslash {\cal I'},  i \ne s,  \\
	&& T_{s} = T_{s}^\ast -t, \label{Cons:2}\\
	&& T_{i} = T_i^\ast,  \forall i \in {\cal I'}.\label{Cons:1}
\end{eqnarray}

For (P2.3.$t$), we can introduce a similar variable transformation as in (P2.2) and solve the resulting problem w.r.t. $\bm x$ (denoted as (P2.4.$t$) and omitted here for brevity) optimally using the interior-point algorithm.
Let $\bm x'_t\triangleq \{x'_{t,i}\}$ denote the optimal solution to (P2.4.$t$).
Similarly to \eqref{Eq:1}, the corresponding reconstructed tile numbers are given by
\begin{eqnarray}\label{Eq:2}
	{\cal T}_t' = \{	T_{i} |T_{i} = \lceil e^{x'_{t,i}} \rceil,  i\in {\cal P}\cup {\cal A}\}.
\end{eqnarray}
Next, we check whether the incumbent total deployment cost, i.e., $c({\cal P},{\cal A},{\cal T}^\ast)$, can be reduced by replacing ${\cal T}^\ast$ with ${\cal T}_t'$, i.e., $c({\cal P},{\cal A}, {\cal T}_t') < c({\cal P},{\cal A}, {\cal T}^\ast)$.
If the above condition is satisfied, we update ${\cal T}^\ast$ as ${\cal T}_t'$ and proceed to solve (P2.3.$t+1$) (as well as (P2.4.$t+1$)).
Otherwise, the successive tile removal terminates, and we update ${\cal I'}$ as ${\cal I'}\cup \{s\}$ and proceed to the tile refinement for the next cell.
It is evident that by applying the above sequential refinement method, the objective value of (P2) or the total deployment cost will be non-increasing with the iterations.

Nonetheless, the ultimate performance of the sequential refinement for all cells in ${\cal P} \cup {\cal A}$ depends critically on the order of cells selected.
In this paper, we propose to determine their selected order based on the following procedures.
First, at the beginning of the sequential refinement, we calculate the increase in the tile numbers of each cell $i, i \in {\cal P} \cup {\cal A}$ after the reconstruction in \eqref{Eq:1}, denoted as $\delta^\ast_i = \lceil e^{x^\ast_{i}} \rceil- e^{x^\ast_{i}}, i\in {\cal P} \cup {\cal A}$.
It is evident that a larger $\delta_i^\ast$ indicates that $e^{x_i^\ast}$ is closer to $\lfloor {e^{x_i^\ast}} \rfloor$ than $\lceil {e^{x_i^\ast}} \rceil$, and thus it may be more likely to reduce the tile number deployed in cell $i$.
As such, we first refine the tile number in cell $s$ with $s=\arg \max_{i\in {\cal P} \cup {\cal A}} \delta^\ast_i$.
Next, in each iteration, if ${\cal T}^\ast$ can be updated as ${\cal T}_t'$ in \eqref{Eq:2}, we update $\delta_i^\ast = \lceil e^{x'_{t,i}} \rceil - e^{x'_{t,i}}, i \in {\cal P} \cup {\cal A}$. 
After the successive tile removal in this iteration terminates, we update $s = \arg\max_{{\cal P} \cup {\cal A} \backslash {\cal I}'} \delta^\ast_i$ and refine the tile number in cell $s$ in the next iteration.
The main procedures of the proposed sequential refinement algorithm are summarized in Algorithm \ref{Algo:tile}.
For Algorithm \ref{Algo:tile}, its computational complexity is mainly due to the use of the Dijkstra algorithm for computing \eqref{Eq:path} and solving the convex problems (P2.2) and (P2.4.$t$), which thus yields polynomial complexity in the order of $|{\cal P}|+|{\cal A}|$.
As a result, Algorithm \ref{Algo:tile} is ensured to incur much lower complexity than full enumeration which requires an exponential complexity in the order of $(T_{0}^{\max})^{ |{\cal P}|+ |{\cal A}|}$.

\subsection{IRS Location Optimization}

In this subsection, we aim to solve the original problem (P1), based on the proposed solution to (P2) via Algorithm \ref{Algo:tile}.
Due to the inherent complex structure of (P1), we enumerate all possible PIRS and AIRS location combinations; while for each PIRS and AIRS location deployment, Algorithm \ref{Algo:tile} can be invoked to obtain the tile number solution.
Then, the solution to (P1) can be obtained as the one that achieves the minimum objective value.
Although such an enumeration involves the search among $\sum_{i = 0}^{|{\cal I}_0|}\sum_{j = 0}^{|{\cal I}_0|-i}\tbinom{|{\cal I}_0|}{i}\tbinom{|{\cal I}_0|-i}{j}$ possible IRS location combinations, most of them can be safely discarded due to the following two reasons.
First, many location combinations cannot yield a feasible solution to (P1).
Second, the total deployment cost by some location combinations may be even higher than the incumbent minimum cost even if we set the tile number per location to its minimum (i.e., ${\cal T} = {\cal T}^{\min}$), especially when $|{\cal P}|$ and/or $|{\cal A}|$ is large.
In both cases, there is no need to implement the sequential refinement process in Algorithm 1.
Hence, only a partial enumeration of all possible location combinations is needed, which incurs considerably lower computational complexity than full enumeration.
Furthermore, (P1) can be optimized offline, and thus the above computational time is practically tolerable.
\begin{figure}[t]
	\centering		
	\includegraphics[scale=0.5]{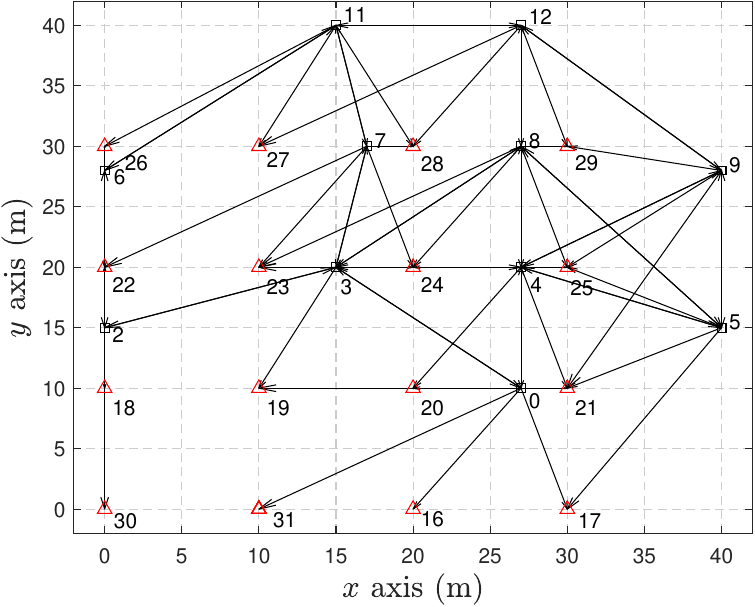}	
	\vspace{-3mm}		
	\caption{LoS graph $G$ of the considered environment.} \label{Fig:Topology1}	
	\vspace{-2mm}
\end{figure}
\begin{figure}[t]
	\centering
	\includegraphics[scale=0.5]{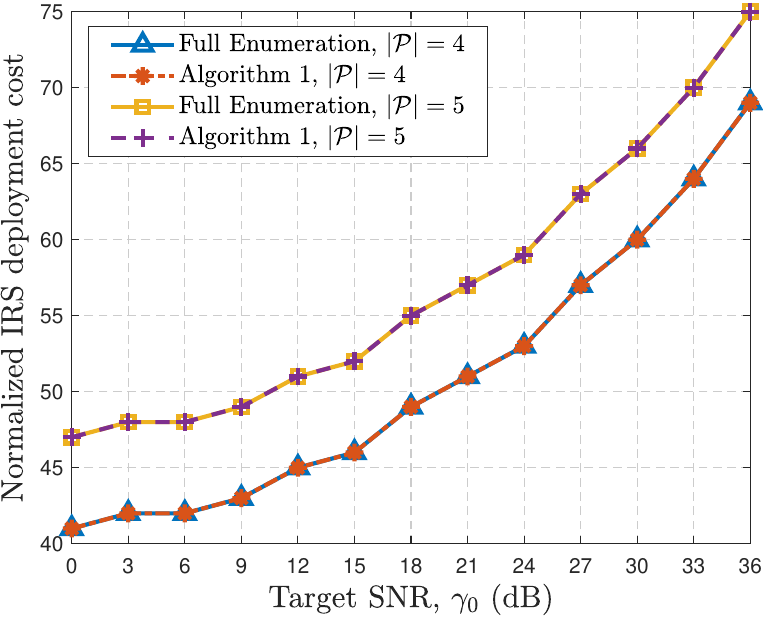}
	\vspace{-2mm}		
	\caption{Total deployment cost versus SNR target with given IRS locations. } \label{Fig:GainSNR_givenloc}
	\vspace{-4mm}
\end{figure}

\begin{figure*}[t]
	\centering
	\vspace{-1mm}
	\subfigure[Joint IRS deployment with $c({\cal P}, {\cal A}, {\cal T})= 46$.]{\includegraphics[scale=0.45]{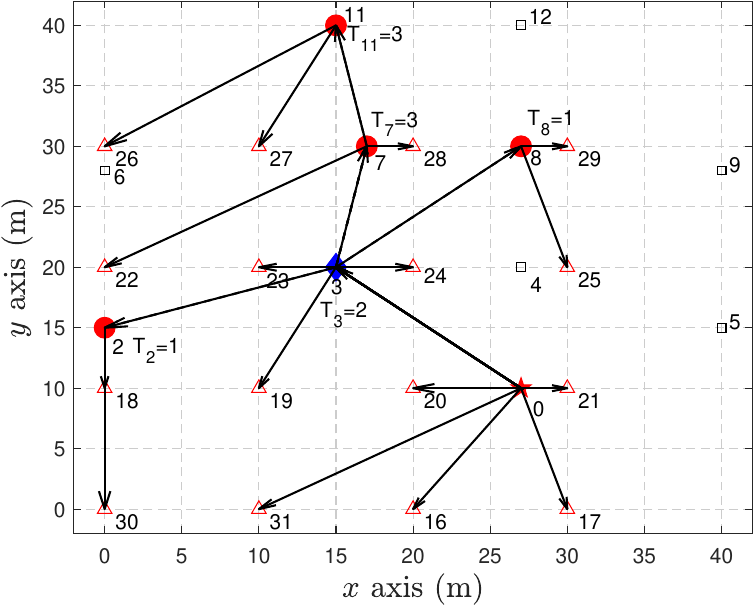}\label{Fig:Joint15}}
	\hspace{-2mm}
	\subfigure[Benchmark 1 with $c({\cal P}, {\cal A}, {\cal T}) = 50$.]{\includegraphics[scale=0.45]{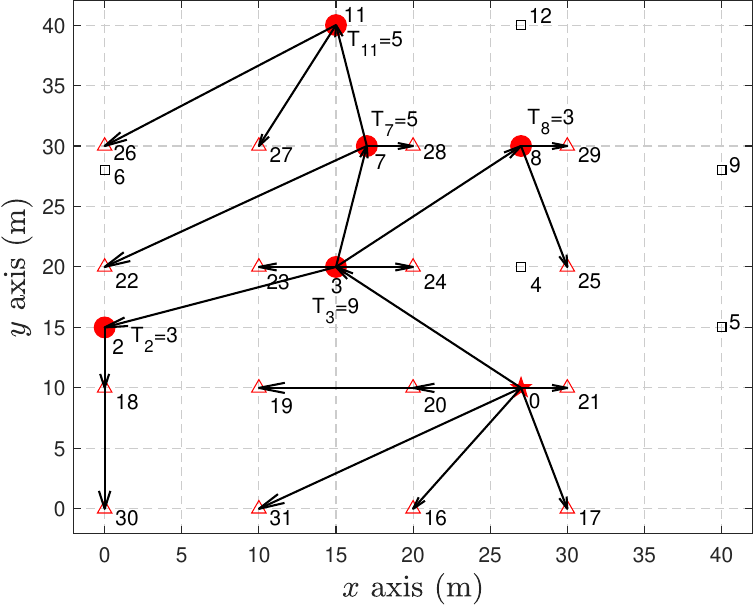}\label{Fig:passive15}}	
	\hspace{-2mm}
	\subfigure[Benchmark 3 with $c({\cal P}, {\cal A}, {\cal T}) =53$.]{\includegraphics[scale=0.45]{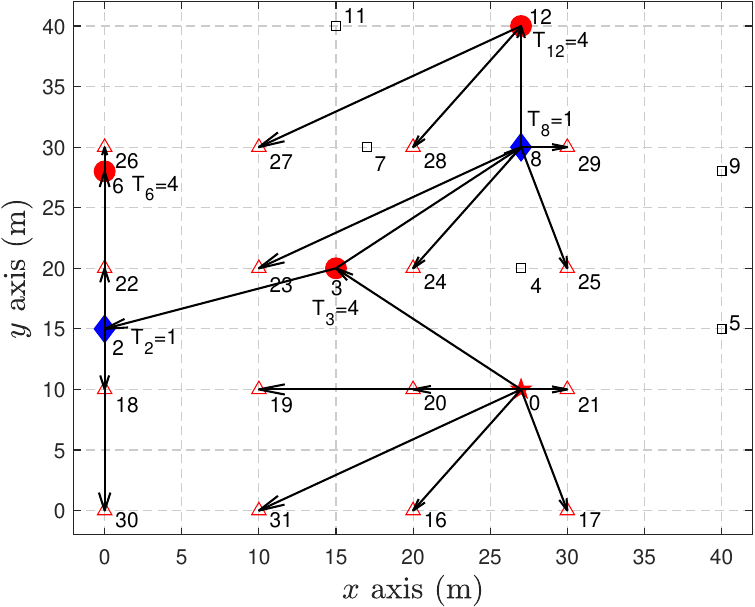}\label{Fig:equal15}}
	\caption{Jointly optimized PIRS/AIRS deployment under different schemes with $\gamma_0 = 15$ dB.}\label{Fig:All-PIRSvsAIRS15}
	\vspace{-0mm}
\end{figure*}

\begin{figure*}[t]
	\centering
	\vspace{-1mm}
	\subfigure[Joint IRS deployment with $c({\cal P}, {\cal A}, {\cal T})=54$.]{\includegraphics[scale=0.45]{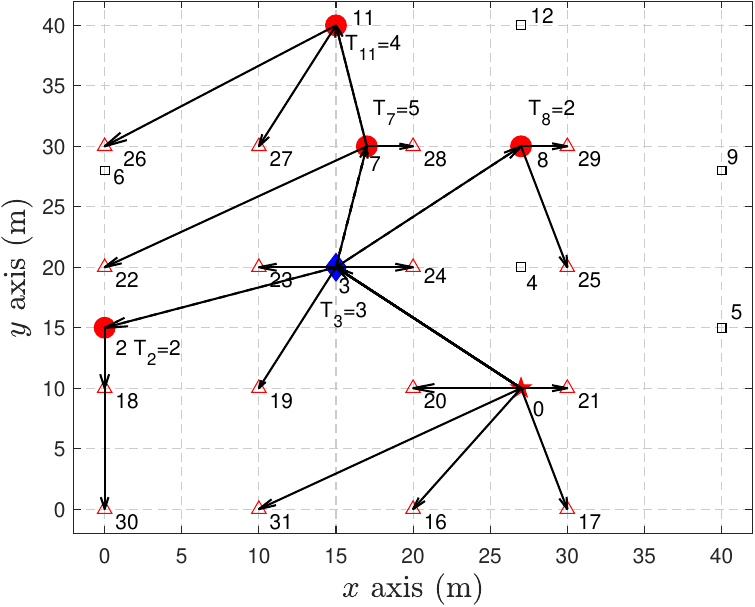}\label{Fig:Joint24}}
	\hspace{-2mm}
	\subfigure[Benchmark 1 with $c({\cal P}, {\cal A}, {\cal T}) = 68$.]{\includegraphics[scale=0.45]{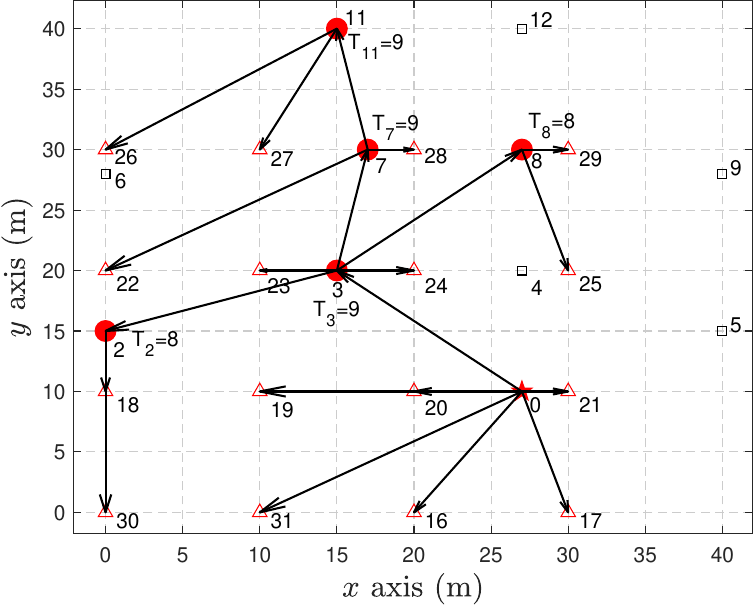}\label{Fig:passive24}}
	\hspace{-2mm}
	\subfigure[Benchmark 3 with $c({\cal P}, {\cal A}, {\cal T}) =63$.]{\includegraphics[scale=0.45]{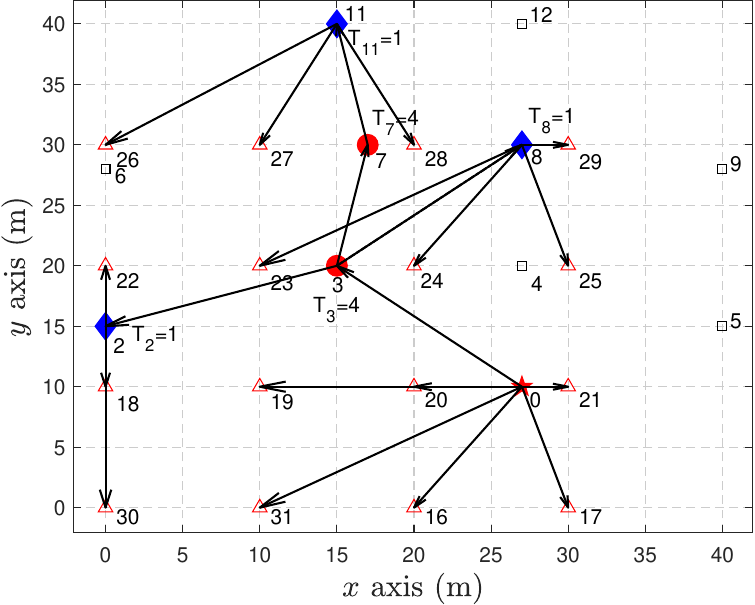}\label{Fig:equal24}}
		\hspace{-2mm}
	\caption{Jointly optimized PIRS/AIRS deployment under different schemes with $\gamma_0 = 25$ dB.}\label{Fig:All-PIRSvsAIRS24}
	\vspace{-4mm}
\end{figure*}

\section{Numerical Results}\label{Sec:Result}
In this section, we provide numerical results to demonstrate the effectiveness of our proposed algorithms for joint PIRS and AIRS deployment optimization.
We consider a typical indoor environment with an area of 40 m $\times$ 40 m that is divided into $J = 16$ cells each with an area of 10 m $\times$ 10 m, among which there exist $I_0 = 10$ candidate locations. 
The BS is assumed to be equipped with $M=10$ antennas and deployed in cell $0$ to cover the maximum number of cells.
In addition, the set of cells containing candidate IRS locations is set to ${\cal I}_0 = \{2,3,4,5,6,7,8,9,11,12\}$.
The LoS graph $G$ of the considered environment is shown in Fig. \ref{Fig:Topology1}, where the nodes representing the candidate IRS locations and the BS are marked by black squares (\textcolor{black}{$\square$}), while the virtual nodes representing all possible user locations in each cell are marked by red triangles (\textcolor{red}{$\triangle$}).
The communication system is assumed to operate at a carrier frequency of 3.5 GHz, with the wavelength $\lambda = 0.087$ m and the reference LoS path gain $\beta_0 = (\lambda/4\pi)^2 = -43$ dB.
The LoS path-loss exponent is set to $\alpha = 2$.
The maximum transmit power of the BS is set to $P_{0} = 30$ dBm, while the maximum amplification power per AIRS reflecting element is set to $P_{A} = -5$ dBm.
The user receiver and AIRS noise power are assumed to be identical to $\sigma^2 = -60$ dBm.
The number of reflecting elements in each passive/active tile's dimension is set to $N = 10$.
The maximum number of tiles that can be deployed at each candidate location is set to $T_{0}^\text {max} = 9$.
The hardware costs per passive and active tile are set to $c_{P} = 1$ and $c_{A} = 3$, respectively, while the cell-use costs for PIRS and AIRS deployment are set to $\alpha_{P,0} = 5$ and $\alpha_{A,0} = 12$, respectively.

In the simulation, to evaluate the performance of the proposed deployment algorithms, we consider the following three benchmarks.
\begin{itemize}
	
	\item {\it Benchmark 1}: All-PIRS deployment with tile number optimization for candidate locations. The performance of this benchmark can be obtained by setting ${\cal A}=\emptyset$ in (P1).

	\item {\it Benchmark 2}: All-PIRS deployment with equal tile number at all candidate locations. The performance of this benchmark can be obtained by setting ${\cal A}=\emptyset$ and $T_{p}=4, \forall p\in {\cal P}$  in (P1).

	\item {\it Benchmark 3}: Joint PIRS and AIRS deployment with equal tile number at all candidate locations. The performance of this benchmark can be obtained by setting $T_{p}=4, \forall p\in {\cal P}$ and  $T_{a} = 1, \forall a\in {\cal A}$ in (P1).
	
\end{itemize}

\subsection{Optimized Tile Numbers with Given IRS Locations}

Fig. \ref{Fig:GainSNR_givenloc} shows the total deployment cost by Algorithm 1 and full enumeration versus the SNR target  $\gamma_{0}$ with given IRS locations.
Here, we consider a single AIRS deployed in cell $3$ (i.e., ${\cal A} = \{3\}$) with $|{\cal A}| = 1$ and multiple PIRSs with the following two deployment designs, i.e., ${\cal P} = \{2,7,8,11\}$ with $|{\cal P}| = 4$ and ${\cal P} = \{2,6,7,8,11\}$ for $|{\cal P}| = 5$.
Accordingly, the total cell-use cost is a constant and equal to $c_{A,0} +4c_{P,0} = 32$ and $c_{A,0} +5c_{P,0}=37$ in the cases of $|{\cal P}| = 4$ and $|{\cal P}| = 5$, respectively.
First, it is observed from Fig. \ref{Fig:GainSNR_givenloc} that the proposed algorithm achieves the same performance as full enumeration in both PIRS deployment designs.
Moreover, in both PIRS deployment designs, the total deployment cost (or the hardware cost) monotonically increases with $\gamma_0$, as expected.
To further manifest the computational efficiency of our proposed Algorithm 1, we compare its running time (in seconds) with that of full enumeration in Table \ref{timetable} under $\gamma_{0} = 36$ dB.
It is observed that the proposed algorithm incurs much shorter computational time than full enumeration while yielding a near-optimal solution to (P2), especially in the case of $|{\cal P}| = 5$, which thus validates the efficiency of Algorithm 1 in solving (P2).

\begin{table}[t]
	\centering
	\vspace{-2mm}
	\caption{Running time (in second) of different algorithms for solving (P2)}\label{timetable}
	\begin{tabular}{|c|c|c|}
		\hline 
		&  $|{\cal P}| = 4$ & $| {\cal P}| = 5$ \\
		\hline
		Full Enumeration & 550.8 & 4235.1 \\
		\hline
		Algorithm 1 &  1.6   & 3.3   \\
		\hline
	\end{tabular}
	\vspace{-4mm}
\end{table}
\subsection{Jointly Optimized PIRS and AIRS Deployment}

Next, we show the performance of our proposed algorithm for joint IRS deployment and tile number optimization.
Figs. \ref{Fig:All-PIRSvsAIRS15} and \ref{Fig:All-PIRSvsAIRS24} show the optimized IRS deployment solutions to (P1) under different values of $\gamma_0$ and by different benchmark schemes.
In both figures, the BS's located cell (cell 0) is marked by a red star (\textcolor{red}{$\star$}), while the cells deployed with PIRSs and AIRSs are marked by red circles (\textcolor{red}{$\bullet$}) and blue diamonds (\textcolor{blue}{$\blacklozenge$}), respectively.
Moreover, the optimized tile number at each selected candidate location (i.e., $T_{i}, i \in {\cal P}\cup{\cal A}$) and the selected path from the BS to each cell to satisfy the SNR target are also shown.

Fig. \ref{Fig:All-PIRSvsAIRS15} shows the optimized IRS deployment solutions under different schemes with $\gamma_0 = 15$ dB.
First, it is observed from Fig. \ref{Fig:Joint15} that our proposed deployment solution deploys one AIRS (with 2 active tiles) and 4 PIRSs (with 8 passive tiles in total) in 5 cells, while leaving the other 5 candidate locations unused.
This implies that the proposed deployment solution can substantially reduce the cell-use cost and thus the total deployment cost.
Second, it is found in Fig. \ref{Fig:Joint15} that the selected cascaded LoS paths from the BS to all cells outside its direct LoS coverage go through the AIRS, which implies its pivotal role in terms of LoS coverage enhancement.
This is because the AIRS is capable of amplifying the incident signal strength and thus is used in most LoS paths to satisfy the SNR target.
Moreover, it is observed from Fig. \ref{Fig:passive15} that Benchmark 1 replaces the AIRS in our proposed deployment solution with a PIRS, which yields a lower cell-use cost than ours.
However, to satisfy the SNR target, Benchmark 1 needs to consume 25 passive tiles in total, which results in a much higher hardware cost than our proposed deployment solution.
As a result, its total deployment cost is higher than ours (50 versus 46).
This observation shows that the joint use of AIRSs and PIRSs may help reduce the total deployment cost by reducing the number of passive tiles thanks to signal strength enhancement by AIRSs, which is in accordance with our analysis in Section III-C.
In addition, it is observed from Fig. \ref{Fig:equal15} that compared to our proposed deployment solution, Benchmark 3 deploys AIRSs in two cells with the same active tiles in total as ours and consumes more passive tiles than ours.
Hence, it incurs both higher hardware cost and cell-use cost than our proposed solution, with the total deployment cost equal to 53, which is even higher than that by Benchmark 1.
This is because Benchmark 3 fails to exploit the design degree of freedom in tile number optimization.

Fig. \ref{Fig:All-PIRSvsAIRS24} compares the optimized deployment solutions by different schemes under $\gamma_0 = 25$ dB.
It is observed from Fig. \ref{Fig:Joint24} that by increasing $\gamma_{0}$ from 15 dB to 25 dB, the optimized IRS locations by the proposed IRS deployment solution keep the same, while the optimized numbers of both active and passive tiles increase to satisfy the more stringent SNR constraints.
As such, the total deployment cost increases from 46 to 54, which validates our discussion in Section III-C.
Even so, its total deployment cost is still smaller than those by benchmarks 1 and 3.
It is also observed from Figs. \ref{Fig:Joint24} and \ref{Fig:passive24} that the increment of passive tile numbers in the proposed deployment solution is much smaller than that in Benchmark 1 (i.e., 5 versus 18) thanks to the use of the AIRS.

\begin{figure}[t]
	\centering
	\includegraphics[scale=0.5]{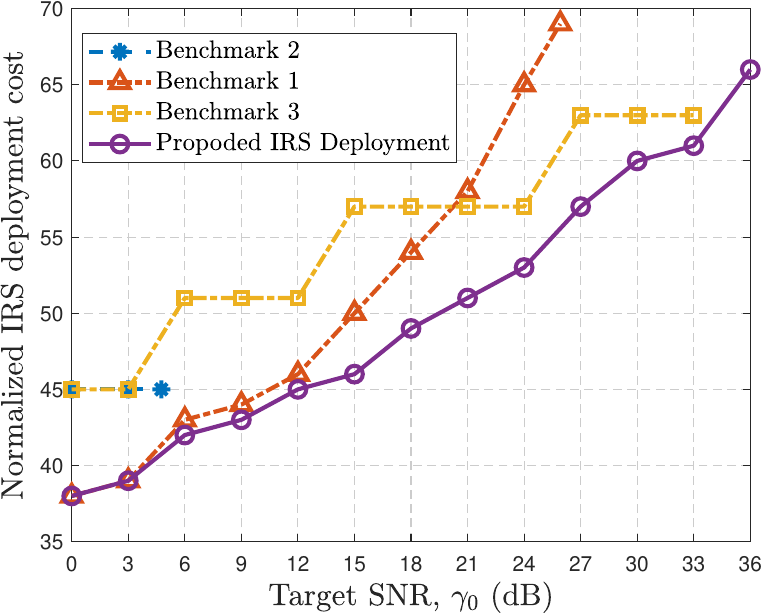}
	\vspace{-2mm}		
	\caption{Total deployment cost versus target SNR. } \label{Fig:SNR-Cost-GainTile}
	\vspace{-6mm}
\end{figure}
\begin{figure}[t]
	\centering
	\includegraphics[scale=0.5]{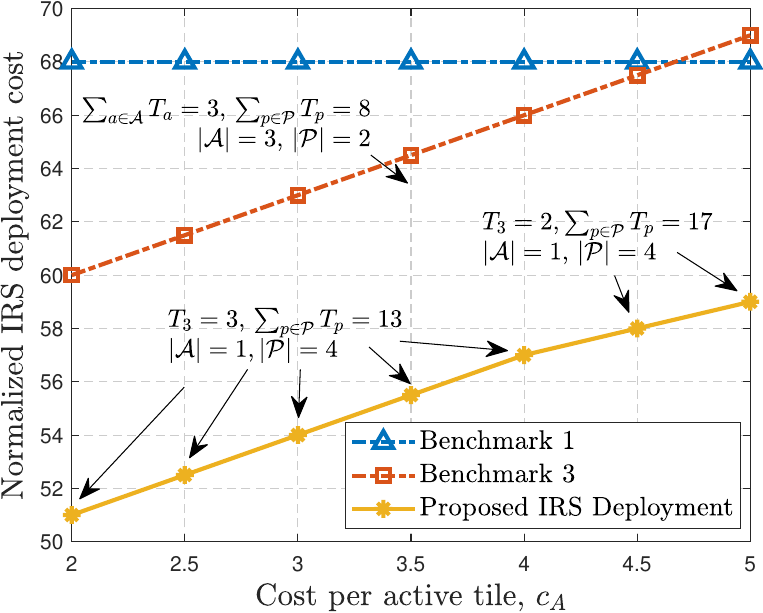}
	\vspace{-2mm}		
	\caption{Total deployment cost versus hardware cost per active tile. } \label{Fig:ration-Cost-GainTile}
	\vspace{-5mm}
\end{figure}

Fig. \ref{Fig:SNR-Cost-GainTile} shows the total deployment costs by different schemes versus the SNR target, $\gamma_{0}$.
First, by comparing the total deployment costs between benchmarks 1 and 2, it is observed that a larger SNR target can be satisfied by the former.
It is also observed that the performance gap between the proposed deployment solution and Benchmark 1 increases with $\gamma_{0}$.
Particularly, the former can satisfy a much larger SNR target than the latter at a much lower cost.
This is because the proposed joint PIRS and AIRS deployment design can significantly reduce the use of passive tiles by deploying AIRSs at several pivotal locations and exploiting their amplification gains, as shown in Figs. \ref{Fig:All-PIRSvsAIRS15} and \ref{Fig:All-PIRSvsAIRS24}.
Furthermore, the proposed deployment solution outperforms Benchmark 3, especially in the low-to-moderate SNR regime by fully exploiting the design degree of freedom in tile number optimization.

Finally, Fig. \ref{Fig:ration-Cost-GainTile} shows the total deployment cost versus the hardware cost per active tile (i.e., $c_{A}$) with $\gamma_0 = 25$ dB.
Note that the performance by Benchmark 2 is not shown in Fig. \ref{Fig:ration-Cost-GainTile} as it fails to satisfy the SNR target with $\gamma_0 = 25$ dB.
It is observed that the total deployment costs by the proposed deployment solution and Benchmark 3 monotonically increase with $c_A$.
Nonetheless, even though $c_{A}$ increases from 2 to 5, the proposed solution considerably outperforms Benchmark 1 without any AIRS.
In addition, it is observed that the proposed deployment solution yields more significant performance gains over Benchmark 3 as $c_{A}$ increases by optimally balancing the tile numbers at different locations.
Accordingly, despite the higher hardware and cell-use costs incurred by AIRS than PIRS, incorporating AIRSs is beneficial to reduce the total deployment cost if their deployment is optimally designed.

\vspace{-1em}

\section{Conclusion}

In this paper, we studied a joint PIRS and AIRS deployment problem to enhance the communication performance in a given region by exploiting multi-PIRS/AIRS reflections.
Based on the proposed graph-based system modeling, it was shown that there exist fundamental trade-offs between minimizing the total deployment cost and maximizing the SNR performance over all cells via the LoS path selection.
To optimally reconcile these trade-offs, we jointly optimized the locations of PIRSs and AIRSs and the tile number at each location.
Such a combinatorial problem was solved by applying a partial enumeration method.
Our numerical results showed that our proposed algorithm achieves near-optimal performance without full enumeration and significantly outperforms other baseline deployment schemes.
It was also shown that the joint use of AIRSs and PIRSs can reduce the total deployment cost by dispensing with a large number of passive reflecting elements for a given SNR target.

This paper can be extended in various promising directions for future work. For example, the user SNRs in this paper are only calculated based on the theoretical LoS channel model. It would be more practical to design the IRS deployment based on actual channel measurements involving non-LoS channels. In addition, how to conduct cell division and find good candidate IRS locations efficiently in practice is also worth investigating in future work.

\vspace{-1em}
\bibliographystyle{IEEEtran}
\bibliography{Ref}

\end{document}